
\catcode`\@=11


\message{Loading jyTeX fonts...}



\font\vptrm=cmr5
\font\vptmit=cmmi5
\font\vptsy=cmsy5
\font\vptbf=cmbx5

\skewchar\vptmit='177 \skewchar\vptsy='60
\fontdimen16 \vptsy=\the\fontdimen17 \vptsy

\def\vpt{\ifmmode\err@badsizechange\else
     \@mathfontinit
     \textfont0=\vptrm  \scriptfont0=\vptrm  \scriptscriptfont0=\vptrm
     \textfont1=\vptmit \scriptfont1=\vptmit \scriptscriptfont1=\vptmit
     \textfont2=\vptsy  \scriptfont2=\vptsy  \scriptscriptfont2=\vptsy
     \textfont3=\xptex  \scriptfont3=\xptex  \scriptscriptfont3=\xptex
     \textfont\bffam=\vptbf
     \scriptfont\bffam=\vptbf
     \scriptscriptfont\bffam=\vptbf
     \@fontstyleinit
     \def\rm{\vptrm\fam=\z@}%
     \def\bf{\vptbf\fam=\bffam}%
     \def\oldstyle{\vptmit\fam=\@ne}%
     \rm\fi}


\font\viptrm=cmr6
\font\viptmit=cmmi6
\font\viptsy=cmsy6
\font\viptbf=cmbx6

\skewchar\viptmit='177 \skewchar\viptsy='60
\fontdimen16 \viptsy=\the\fontdimen17 \viptsy

\def\vipt{\ifmmode\err@badsizechange\else
     \@mathfontinit
     \textfont0=\viptrm  \scriptfont0=\vptrm  \scriptscriptfont0=\vptrm
     \textfont1=\viptmit \scriptfont1=\vptmit \scriptscriptfont1=\vptmit
     \textfont2=\viptsy  \scriptfont2=\vptsy  \scriptscriptfont2=\vptsy
     \textfont3=\xptex   \scriptfont3=\xptex  \scriptscriptfont3=\xptex
     \textfont\bffam=\viptbf
     \scriptfont\bffam=\vptbf
     \scriptscriptfont\bffam=\vptbf
     \@fontstyleinit
     \def\rm{\viptrm\fam=\z@}%
     \def\bf{\viptbf\fam=\bffam}%
     \def\oldstyle{\viptmit\fam=\@ne}%
     \rm\fi}


\font\viiptrm=cmr7
\font\viiptmit=cmmi7
\font\viiptsy=cmsy7
\font\viiptit=cmti7
\font\viiptbf=cmbx7

\skewchar\viiptmit='177 \skewchar\viiptsy='60
\fontdimen16 \viiptsy=\the\fontdimen17 \viiptsy

\def\viipt{\ifmmode\err@badsizechange\else
     \@mathfontinit
     \textfont0=\viiptrm  \scriptfont0=\vptrm  \scriptscriptfont0=\vptrm
     \textfont1=\viiptmit \scriptfont1=\vptmit \scriptscriptfont1=\vptmit
     \textfont2=\viiptsy  \scriptfont2=\vptsy  \scriptscriptfont2=\vptsy
     \textfont3=\xptex    \scriptfont3=\xptex  \scriptscriptfont3=\xptex
     \textfont\itfam=\viiptit
     \scriptfont\itfam=\viiptit
     \scriptscriptfont\itfam=\viiptit
     \textfont\bffam=\viiptbf
     \scriptfont\bffam=\vptbf
     \scriptscriptfont\bffam=\vptbf
     \@fontstyleinit
     \def\rm{\viiptrm\fam=\z@}%
     \def\it{\viiptit\fam=\itfam}%
     \def\bf{\viiptbf\fam=\bffam}%
     \def\oldstyle{\viiptmit\fam=\@ne}%
     \rm\fi}


\font\viiiptrm=cmr8
\font\viiiptmit=cmmi8
\font\viiiptsy=cmsy8
\font\viiiptit=cmti8
\font\viiiptbf=cmbx8

\skewchar\viiiptmit='177 \skewchar\viiiptsy='60
\fontdimen16 \viiiptsy=\the\fontdimen17 \viiiptsy

\def\viiipt{\ifmmode\err@badsizechange\else
     \@mathfontinit
     \textfont0=\viiiptrm  \scriptfont0=\viptrm  \scriptscriptfont0=\vptrm
     \textfont1=\viiiptmit \scriptfont1=\viptmit \scriptscriptfont1=\vptmit
     \textfont2=\viiiptsy  \scriptfont2=\viptsy  \scriptscriptfont2=\vptsy
     \textfont3=\xptex     \scriptfont3=\xptex   \scriptscriptfont3=\xptex
     \textfont\itfam=\viiiptit
     \scriptfont\itfam=\viiptit
     \scriptscriptfont\itfam=\viiptit
     \textfont\bffam=\viiiptbf
     \scriptfont\bffam=\viptbf
     \scriptscriptfont\bffam=\vptbf
     \@fontstyleinit
     \def\rm{\viiiptrm\fam=\z@}%
     \def\it{\viiiptit\fam=\itfam}%
     \def\bf{\viiiptbf\fam=\bffam}%
     \def\oldstyle{\viiiptmit\fam=\@ne}%
     \rm\fi}


\def\getixpt{%
     \font\ixptrm=cmr9
     \font\ixptmit=cmmi9
     \font\ixptsy=cmsy9
     \font\ixptit=cmti9
     \font\ixptbf=cmbx9
     \skewchar\ixptmit='177 \skewchar\ixptsy='60
     \fontdimen16 \ixptsy=\the\fontdimen17 \ixptsy}

\def\ixpt{\ifmmode\err@badsizechange\else
     \@mathfontinit
     \textfont0=\ixptrm  \scriptfont0=\viiptrm  \scriptscriptfont0=\vptrm
     \textfont1=\ixptmit \scriptfont1=\viiptmit \scriptscriptfont1=\vptmit
     \textfont2=\ixptsy  \scriptfont2=\viiptsy  \scriptscriptfont2=\vptsy
     \textfont3=\xptex   \scriptfont3=\xptex    \scriptscriptfont3=\xptex
     \textfont\itfam=\ixptit
     \scriptfont\itfam=\viiptit
     \scriptscriptfont\itfam=\viiptit
     \textfont\bffam=\ixptbf
     \scriptfont\bffam=\viiptbf
     \scriptscriptfont\bffam=\vptbf
     \@fontstyleinit
     \def\rm{\ixptrm\fam=\z@}%
     \def\it{\ixptit\fam=\itfam}%
     \def\bf{\ixptbf\fam=\bffam}%
     \def\oldstyle{\ixptmit\fam=\@ne}%
     \rm\fi}


\font\xptrm=cmr10
\font\xptmit=cmmi10
\font\xptsy=cmsy10
\font\xptex=cmex10
\font\xptit=cmti10
\font\xptsl=cmsl10
\font\xptbf=cmbx10
\font\xpttt=cmtt10
\font\xptss=cmss10
\font\xptsc=cmcsc10
\font\xptbfs=cmb10
\font\xptbmit=cmmib10

\skewchar\xptmit='177 \skewchar\xptbmit='177 \skewchar\xptsy='60
\fontdimen16 \xptsy=\the\fontdimen17 \xptsy

\def\xpt{\ifmmode\err@badsizechange\else
     \@mathfontinit
     \textfont0=\xptrm  \scriptfont0=\viiptrm  \scriptscriptfont0=\vptrm
     \textfont1=\xptmit \scriptfont1=\viiptmit \scriptscriptfont1=\vptmit
     \textfont2=\xptsy  \scriptfont2=\viiptsy  \scriptscriptfont2=\vptsy
     \textfont3=\xptex  \scriptfont3=\xptex    \scriptscriptfont3=\xptex
     \textfont\itfam=\xptit
     \scriptfont\itfam=\viiptit
     \scriptscriptfont\itfam=\viiptit
     \textfont\bffam=\xptbf
     \scriptfont\bffam=\viiptbf
     \scriptscriptfont\bffam=\vptbf
     \textfont\bfsfam=\xptbfs
     \scriptfont\bfsfam=\viiptbf
     \scriptscriptfont\bfsfam=\vptbf
     \textfont\bmitfam=\xptbmit
     \scriptfont\bmitfam=\viiptmit
     \scriptscriptfont\bmitfam=\vptmit
     \@fontstyleinit
     \def\rm{\xptrm\fam=\z@}%
     \def\it{\xptit\fam=\itfam}%
     \def\sl{\xptsl}%
     \def\bf{\xptbf\fam=\bffam}%
     \def\tt{\xpttt}%
     \def\ss{\xptss}%
     \def\sc{\xptsc}%
     \def\bfs{\xptbfs\fam=\bfsfam}%
     \def\bmit{\fam=\bmitfam}%
     \def\oldstyle{\xptmit\fam=\@ne}%
     \rm\fi}


\def\getxipt{%
     \font\xiptrm=cmr10  scaled\magstephalf
     \font\xiptmit=cmmi10 scaled\magstephalf
     \font\xiptsy=cmsy10 scaled\magstephalf
     \font\xiptex=cmex10 scaled\magstephalf
     \font\xiptit=cmti10 scaled\magstephalf
     \font\xiptsl=cmsl10 scaled\magstephalf
     \font\xiptbf=cmbx10 scaled\magstephalf
     \font\xipttt=cmtt10 scaled\magstephalf
     \font\xiptss=cmss10 scaled\magstephalf
     \skewchar\xiptmit='177 \skewchar\xiptsy='60
     \fontdimen16 \xiptsy=\the\fontdimen17 \xiptsy}

\def\xipt{\ifmmode\err@badsizechange\else
     \@mathfontinit
     \textfont0=\xiptrm  \scriptfont0=\viiiptrm  \scriptscriptfont0=\viptrm
     \textfont1=\xiptmit \scriptfont1=\viiiptmit \scriptscriptfont1=\viptmit
     \textfont2=\xiptsy  \scriptfont2=\viiiptsy  \scriptscriptfont2=\viptsy
     \textfont3=\xiptex  \scriptfont3=\xptex     \scriptscriptfont3=\xptex
     \textfont\itfam=\xiptit
     \scriptfont\itfam=\viiiptit
     \scriptscriptfont\itfam=\viiptit
     \textfont\bffam=\xiptbf
     \scriptfont\bffam=\viiiptbf
     \scriptscriptfont\bffam=\viptbf
     \@fontstyleinit
     \def\rm{\xiptrm\fam=\z@}%
     \def\it{\xiptit\fam=\itfam}%
     \def\sl{\xiptsl}%
     \def\bf{\xiptbf\fam=\bffam}%
     \def\tt{\xipttt}%
     \def\ss{\xiptss}%
     \def\oldstyle{\xiptmit\fam=\@ne}%
     \rm\fi}


\font\xiiptrm=cmr12
\font\xiiptmit=cmmi12
\font\xiiptsy=cmsy10  scaled\magstep1
\font\xiiptex=cmex10  scaled\magstep1
\font\xiiptit=cmti12
\font\xiiptsl=cmsl12
\font\xiiptbf=cmbx12
\font\xiipttt=cmtt12
\font\xiiptss=cmss12
\font\xiiptsc=cmcsc10 scaled\magstep1
\font\xiiptbfs=cmb10  scaled\magstep1
\font\xiiptbmit=cmmib10 scaled\magstep1

\skewchar\xiiptmit='177 \skewchar\xiiptbmit='177 \skewchar\xiiptsy='60
\fontdimen16 \xiiptsy=\the\fontdimen17 \xiiptsy

\def\xiipt{\ifmmode\err@badsizechange\else
     \@mathfontinit
     \textfont0=\xiiptrm  \scriptfont0=\viiiptrm  \scriptscriptfont0=\viptrm
     \textfont1=\xiiptmit \scriptfont1=\viiiptmit \scriptscriptfont1=\viptmit
     \textfont2=\xiiptsy  \scriptfont2=\viiiptsy  \scriptscriptfont2=\viptsy
     \textfont3=\xiiptex  \scriptfont3=\xptex     \scriptscriptfont3=\xptex
     \textfont\itfam=\xiiptit
     \scriptfont\itfam=\viiiptit
     \scriptscriptfont\itfam=\viiptit
     \textfont\bffam=\xiiptbf
     \scriptfont\bffam=\viiiptbf
     \scriptscriptfont\bffam=\viptbf
     \textfont\bfsfam=\xiiptbfs
     \scriptfont\bfsfam=\viiiptbf
     \scriptscriptfont\bfsfam=\viptbf
     \textfont\bmitfam=\xiiptbmit
     \scriptfont\bmitfam=\viiiptmit
     \scriptscriptfont\bmitfam=\viptmit
     \@fontstyleinit
     \def\rm{\xiiptrm\fam=\z@}%
     \def\it{\xiiptit\fam=\itfam}%
     \def\sl{\xiiptsl}%
     \def\bf{\xiiptbf\fam=\bffam}%
     \def\tt{\xiipttt}%
     \def\ss{\xiiptss}%
     \def\sc{\xiiptsc}%
     \def\bfs{\xiiptbfs\fam=\bfsfam}%
     \def\bmit{\fam=\bmitfam}%
     \def\oldstyle{\xiiptmit\fam=\@ne}%
     \rm\fi}


\def\getxiiipt{%
     \font\xiiiptrm=cmr12  scaled\magstephalf
     \font\xiiiptmit=cmmi12 scaled\magstephalf
     \font\xiiiptsy=cmsy9  scaled\magstep2
     \font\xiiiptit=cmti12 scaled\magstephalf
     \font\xiiiptsl=cmsl12 scaled\magstephalf
     \font\xiiiptbf=cmbx12 scaled\magstephalf
     \font\xiiipttt=cmtt12 scaled\magstephalf
     \font\xiiiptss=cmss12 scaled\magstephalf
     \skewchar\xiiiptmit='177 \skewchar\xiiiptsy='60
     \fontdimen16 \xiiiptsy=\the\fontdimen17 \xiiiptsy}

\def\xiiipt{\ifmmode\err@badsizechange\else
     \@mathfontinit
     \textfont0=\xiiiptrm  \scriptfont0=\xptrm  \scriptscriptfont0=\viiptrm
     \textfont1=\xiiiptmit \scriptfont1=\xptmit \scriptscriptfont1=\viiptmit
     \textfont2=\xiiiptsy  \scriptfont2=\xptsy  \scriptscriptfont2=\viiptsy
     \textfont3=\xivptex   \scriptfont3=\xptex  \scriptscriptfont3=\xptex
     \textfont\itfam=\xiiiptit
     \scriptfont\itfam=\xptit
     \scriptscriptfont\itfam=\viiptit
     \textfont\bffam=\xiiiptbf
     \scriptfont\bffam=\xptbf
     \scriptscriptfont\bffam=\viiptbf
     \@fontstyleinit
     \def\rm{\xiiiptrm\fam=\z@}%
     \def\it{\xiiiptit\fam=\itfam}%
     \def\sl{\xiiiptsl}%
     \def\bf{\xiiiptbf\fam=\bffam}%
     \def\tt{\xiiipttt}%
     \def\ss{\xiiiptss}%
     \def\oldstyle{\xiiiptmit\fam=\@ne}%
     \rm\fi}


\font\xivptrm=cmr12   scaled\magstep1
\font\xivptmit=cmmi12  scaled\magstep1
\font\xivptsy=cmsy10  scaled\magstep2
\font\xivptex=cmex10  scaled\magstep2
\font\xivptit=cmti12  scaled\magstep1
\font\xivptsl=cmsl12  scaled\magstep1
\font\xivptbf=cmbx12  scaled\magstep1
\font\xivpttt=cmtt12  scaled\magstep1
\font\xivptss=cmss12  scaled\magstep1
\font\xivptsc=cmcsc10 scaled\magstep2
\font\xivptbfs=cmb10  scaled\magstep2
\font\xivptbmit=cmmib10 scaled\magstep2

\skewchar\xivptmit='177 \skewchar\xivptbmit='177 \skewchar\xivptsy='60
\fontdimen16 \xivptsy=\the\fontdimen17 \xivptsy

\def\xivpt{\ifmmode\err@badsizechange\else
     \@mathfontinit
     \textfont0=\xivptrm  \scriptfont0=\xptrm  \scriptscriptfont0=\viiptrm
     \textfont1=\xivptmit \scriptfont1=\xptmit \scriptscriptfont1=\viiptmit
     \textfont2=\xivptsy  \scriptfont2=\xptsy  \scriptscriptfont2=\viiptsy
     \textfont3=\xivptex  \scriptfont3=\xptex  \scriptscriptfont3=\xptex
     \textfont\itfam=\xivptit
     \scriptfont\itfam=\xptit
     \scriptscriptfont\itfam=\viiptit
     \textfont\bffam=\xivptbf
     \scriptfont\bffam=\xptbf
     \scriptscriptfont\bffam=\viiptbf
     \textfont\bfsfam=\xivptbfs
     \scriptfont\bfsfam=\xptbfs
     \scriptscriptfont\bfsfam=\viiptbf
     \textfont\bmitfam=\xivptbmit
     \scriptfont\bmitfam=\xptbmit
     \scriptscriptfont\bmitfam=\viiptmit
     \@fontstyleinit
     \def\rm{\xivptrm\fam=\z@}%
     \def\it{\xivptit\fam=\itfam}%
     \def\sl{\xivptsl}%
     \def\bf{\xivptbf\fam=\bffam}%
     \def\tt{\xivpttt}%
     \def\ss{\xivptss}%
     \def\sc{\xivptsc}%
     \def\bfs{\xivptbfs\fam=\bfsfam}%
     \def\bmit{\fam=\bmitfam}%
     \def\oldstyle{\xivptmit\fam=\@ne}%
     \rm\fi}


\font\xviiptrm=cmr17
\font\xviiptmit=cmmi12 scaled\magstep2
\font\xviiptsy=cmsy10 scaled\magstep3
\font\xviiptex=cmex10 scaled\magstep3
\font\xviiptit=cmti12 scaled\magstep2
\font\xviiptbf=cmbx12 scaled\magstep2
\font\xviiptbfs=cmb10 scaled\magstep3

\skewchar\xviiptmit='177 \skewchar\xviiptsy='60
\fontdimen16 \xviiptsy=\the\fontdimen17 \xviiptsy

\def\xviipt{\ifmmode\err@badsizechange\else
     \@mathfontinit
     \textfont0=\xviiptrm  \scriptfont0=\xiiptrm  \scriptscriptfont0=\viiiptrm
     \textfont1=\xviiptmit \scriptfont1=\xiiptmit \scriptscriptfont1=\viiiptmit
     \textfont2=\xviiptsy  \scriptfont2=\xiiptsy  \scriptscriptfont2=\viiiptsy
     \textfont3=\xviiptex  \scriptfont3=\xiiptex  \scriptscriptfont3=\xptex
     \textfont\itfam=\xviiptit
     \scriptfont\itfam=\xiiptit
     \scriptscriptfont\itfam=\viiiptit
     \textfont\bffam=\xviiptbf
     \scriptfont\bffam=\xiiptbf
     \scriptscriptfont\bffam=\viiiptbf
     \textfont\bfsfam=\xviiptbfs
     \scriptfont\bfsfam=\xiiptbfs
     \scriptscriptfont\bfsfam=\viiiptbf
     \@fontstyleinit
     \def\rm{\xviiptrm\fam=\z@}%
     \def\it{\xviiptit\fam=\itfam}%
     \def\bf{\xviiptbf\fam=\bffam}%
     \def\bfs{\xviiptbfs\fam=\bfsfam}%
     \def\oldstyle{\xviiptmit\fam=\@ne}%
     \rm\fi}


\font\xxiptrm=cmr17  scaled\magstep1


\def\xxipt{\ifmmode\err@badsizechange\else
     \@mathfontinit
     \@fontstyleinit
     \def\rm{\xxiptrm\fam=\z@}%
     \rm\fi}


\font\xxvptrm=cmr17  scaled\magstep2


\def\xxvpt{\ifmmode\err@badsizechange\else
     \@mathfontinit
     \@fontstyleinit
     \def\rm{\xxvptrm\fam=\z@}%
     \rm\fi}




\message{Loading jyTeX macros...}

\message{modifications to plain.tex,}


\def\newcount{\alloc@0\count\countdef\insc@unt}
\def\newdimen{\alloc@1\dimen\dimendef\insc@unt}
\def\newskip{\alloc@2\skip\skipdef\insc@unt}
\def\newmuskip{\alloc@3\muskip\muskipdef\@cclvi}
\def\newbox{\alloc@4\box\chardef\insc@unt}
\def\newtoks{\alloc@5\toks\toksdef\@cclvi}
\def\newhelp#1#2{\newtoks#1\global#1\expandafter{\csname#2\endcsname}}
\def\newread{\alloc@6\read\chardef\sixt@@n}
\def\newwrite{\alloc@7\write\chardef\sixt@@n}
\def\newfam{\alloc@8\fam\chardef\sixt@@n}
\def\newinsert#1{\global\advance\insc@unt by\m@ne
     \ch@ck0\insc@unt\count
     \ch@ck1\insc@unt\dimen
     \ch@ck2\insc@unt\skip
     \ch@ck4\insc@unt\box
     \allocationnumber=\insc@unt
     \global\chardef#1=\allocationnumber
     \wlog{\string#1=\string\insert\the\allocationnumber}}
\def\newif#1{\count@\escapechar \escapechar\m@ne
     \expandafter\expandafter\expandafter
          \xdef\@if#1{true}{\let\noexpand#1=\noexpand\iftrue}%
     \expandafter\expandafter\expandafter
          \xdef\@if#1{false}{\let\noexpand#1=\noexpand\iffalse}%
     \global\@if#1{false}\escapechar=\count@}


\newlinechar=`\^^J
\overfullrule=0pt




\let\itfam=\undefined

\let\bffam=\undefined

\count18=3


\chardef\sharps="19


\mathchardef\alpha="710B
\mathchardef\beta="710C
\mathchardef\gamma="710D
\mathchardef\delta="710E
\mathchardef\epsilon="710F
\mathchardef\zeta="7110
\mathchardef\eta="7111
\mathchardef\theta="7112
\mathchardef\iota="7113
\mathchardef\kappa="7114
\mathchardef\lambda="7115
\mathchardef\mu="7116
\mathchardef\nu="7117
\mathchardef\xi="7118
\mathchardef\pi="7119
\mathchardef\rho="711A
\mathchardef\sigma="711B
\mathchardef\tau="711C
\mathchardef\upsilon="711D
\mathchardef\phi="711E
\mathchardef\chi="711F
\mathchardef\psi="7120
\mathchardef\omega="7121
\mathchardef\varepsilon="7122
\mathchardef\vartheta="7123
\mathchardef\varpi="7124
\mathchardef\varrho="7125
\mathchardef\varsigma="7126
\mathchardef\varphi="7127
\mathchardef\imath="717B
\mathchardef\jmath="717C
\mathchardef\ell="7160
\mathchardef\wp="717D
\mathchardef\partial="7140
\mathchardef\flat="715B
\mathchardef\natural="715C
\mathchardef\sharp="715D



\def\angle{{\vbox{\ialign{$\m@th\scriptstyle##$\crcr
     \not\mathrel{\mkern14mu}\crcr
     \noalign{\nointerlineskip}
     \mkern2.5mu\leaders\hrule height.34\rp@\hfill\mkern2.5mu\crcr}}}}
\def\vdots{\vbox{\baselineskip4\rp@ \lineskiplimit\z@
     \kern6\rp@\hbox{.}\hbox{.}\hbox{.}}}
\def\ddots{\mathinner{\mkern1mu\raise7\rp@\vbox{\kern7\rp@\hbox{.}}\mkern2mu
     \raise4\rp@\hbox{.}\mkern2mu\raise\rp@\hbox{.}\mkern1mu}}
\def\overrightarrow#1{\vbox{\ialign{##\crcr
     \rightarrowfill\crcr
     \noalign{\kern-\rp@\nointerlineskip}
     $\hfil\displaystyle{#1}\hfil$\crcr}}}
\def\overleftarrow#1{\vbox{\ialign{##\crcr
     \leftarrowfill\crcr
     \noalign{\kern-\rp@\nointerlineskip}
     $\hfil\displaystyle{#1}\hfil$\crcr}}}
\def\overbrace#1{\mathop{\vbox{\ialign{##\crcr
     \noalign{\kern3\rp@}
     \downbracefill\crcr
     \noalign{\kern3\rp@\nointerlineskip}
     $\hfil\displaystyle{#1}\hfil$\crcr}}}\limits}
\def\underbrace#1{\mathop{\vtop{\ialign{##\crcr
     $\hfil\displaystyle{#1}\hfil$\crcr
     \noalign{\kern3\rp@\nointerlineskip}
     \upbracefill\crcr
     \noalign{\kern3\rp@}}}}\limits}
\def\big#1{{\hbox{$\left#1\vbox to8.5\rp@ {}\right.\n@space$}}}
\def\Big#1{{\hbox{$\left#1\vbox to11.5\rp@ {}\right.\n@space$}}}
\def\bigg#1{{\hbox{$\left#1\vbox to14.5\rp@ {}\right.\n@space$}}}
\def\Bigg#1{{\hbox{$\left#1\vbox to17.5\rp@ {}\right.\n@space$}}}
\def\@vereq#1#2{\lower.5\rp@\vbox{\baselineskip\z@skip\lineskip-.5\rp@
     \ialign{$\m@th#1\hfil##\hfil$\crcr#2\crcr=\crcr}}}
\def\rlh@#1{\vcenter{\hbox{\ooalign{\raise2\rp@
     \hbox{$#1\rightharpoonup$}\crcr
     $#1\leftharpoondown$}}}}
\def\bordermatrix#1{\begingroup\m@th
     \setbox\z@\vbox{%
          \def\cr{\crcr\noalign{\kern2\rp@\global\let\cr\endline}}%
          \ialign{$##$\hfil\kern2\rp@\kern\p@renwd
               &\thinspace\hfil$##$\hfil&&\quad\hfil$##$\hfil\crcr
               \omit\strut\hfil\crcr
               \noalign{\kern-\baselineskip}%
               #1\crcr\omit\strut\cr}}%
     \setbox\tw@\vbox{\unvcopy\z@\global\setbox\@ne\lastbox}%
     \setbox\tw@\hbox{\unhbox\@ne\unskip\global\setbox\@ne\lastbox}%
     \setbox\tw@\hbox{$\kern\wd\@ne\kern-\p@renwd\left(\kern-\wd\@ne
          \global\setbox\@ne\vbox{\box\@ne\kern2\rp@}%
          \vcenter{\kern-\ht\@ne\unvbox\z@\kern-\baselineskip}%
          \,\right)$}%
     \null\;\vbox{\kern\ht\@ne\box\tw@}\endgroup}
\def\endinsert{\egroup
     \if@mid\dimen@\ht\z@
          \advance\dimen@\dp\z@
          \advance\dimen@12\rp@
          \advance\dimen@\pagetotal
          \ifdim\dimen@>\pagegoal\@midfalse\p@gefalse\fi
     \fi
     \if@mid\bigskip\box\z@
          \bigbreak
     \else\insert\topins{\penalty100 \splittopskip\z@skip
               \splitmaxdepth\maxdimen\floatingpenalty\z@
               \ifp@ge\dimen@\dp\z@
                    \vbox to\vsize{\unvbox\z@\kern-\dimen@}%
               \else\box\z@\nobreak\bigskip
               \fi}%
     \fi
     \endgroup}


\def\cases#1{\left\{\,\vcenter{\m@th
     \ialign{$##\hfil$&\quad##\hfil\crcr#1\crcr}}\right.}
\def\matrix#1{\null\,\vcenter{\m@th
     \ialign{\hfil$##$\hfil&&\quad\hfil$##$\hfil\crcr
          \mathstrut\crcr
          \noalign{\kern-\baselineskip}
          #1\crcr
          \mathstrut\crcr
          \noalign{\kern-\baselineskip}}}\,}


\newif\ifraggedbottom

\def\raggedbottom{\ifraggedbottom\else
     \advance\topskip by\z@ plus60pt \raggedbottomtrue\fi}%
\def\normalbottom{\ifraggedbottom
     \advance\topskip by\z@ plus-60pt \raggedbottomfalse\fi}

\message{hacks,}


\toksdef\toks@i=1
\toksdef\toks@ii=2


\def\TeX{T\kern-.1667em \lower.5ex \hbox{E}\kern-.125em X\null}
\def\jyTeX{{\leavevmode
     \raise.587ex \hbox{\it\j}\kern-.1em \lower.048ex \hbox{\it y}\kern-.12em
     \TeX}}

\let\then=\iftrue
\def\ifnoarg#1\then{\def\hack@{#1}\ifx\hack@\empty}
\def\ifundefined#1\then{%
     \expandafter\ifx\csname\expandafter\blank\string#1\endcsname\relax}
\def\useif#1\then{\csname#1\endcsname}
\def\usename#1{\csname#1\endcsname}
\def\useafter#1#2{\expandafter#1\csname#2\endcsname}

\long\def\loop#1\repeat{\def\@iterate{#1\expandafter\@iterate\fi}\@iterate
     \let\@iterate=\relax}

\let\TeXend=\end
\def\begin#1{\begingroup\def\@@blockname{#1}\usename{begin#1}}
\def\end#1{\usename{end#1}\def\hack@{#1}%
     \ifx\@@blockname\hack@
          \endgroup
     \else\err@badgroup\hack@\@@blockname
     \fi}
\def\@@blockname{}

\def\defaultoption[#1]#2{%
     \def\hack@{\ifx\hack@ii[\toks@={#2}\else\toks@={#2[#1]}\fi\the\toks@}%
     \futurelet\hack@ii\hack@}

\def\markup#1{\let\@@marksf=\empty
     \ifhmode\edef\@@marksf{\spacefactor=\the\spacefactor\relax}\/\fi
     ${}^{\hbox{\subscriptfonts#1}}$\@@marksf}


\newtoks\shortyear
\newtoks\militaryhour
\newtoks\standardhour
\newtoks\minute
\newtoks\amorpm

\def\settime{\count@=\time\divide\count@ by60
     \militaryhour=\expandafter{\number\count@}%
     {\multiply\count@ by-60 \advance\count@ by\time
          \xdef\hack@{\ifnum\count@<10 0\fi\number\count@}}%
     \minute=\expandafter{\hack@}%
     \ifnum\count@<12
          \amorpm={am}
     \else\amorpm={pm}
          \ifnum\count@>12 \advance\count@ by-12 \fi
     \fi
     \standardhour=\expandafter{\number\count@}%
     \def\hack@19##1##2{\shortyear={##1##2}}%
          \expandafter\hack@\the\year}

\def\monthword#1{%
     \ifcase#1
          $\bullet$\err@badcountervalue{monthword}%
          \or January\or February\or March\or April\or May\or June%
          \or July\or August\or September\or October\or November\or December%
     \else$\bullet$\err@badcountervalue{monthword}%
     \fi}

\def\monthabbr#1{%
     \ifcase#1
          $\bullet$\err@badcountervalue{monthabbr}%
          \or Jan\or Feb\or Mar\or Apr\or May\or Jun%
          \or Jul\or Aug\or Sep\or Oct\or Nov\or Dec%
     \else$\bullet$\err@badcountervalue{monthabbr}%
     \fi}

\def\militarytime{\the\militaryhour:\the\minute}
\def\standardtime{\the\standardhour:\the\minute}


\def\@setnumstyle#1#2{\expandafter\global\expandafter\expandafter
     \expandafter\let\expandafter\expandafter
     \csname @\expandafter\blank\string#1style\endcsname
     \csname#2\endcsname}
\def\numstyle#1{\usename{@\expandafter\blank\string#1style}#1}
\def\ifblank#1\then{\useafter\ifx{@\expandafter\blank\string#1}\blank}

\def\blank#1{}

\def\Roman#1{\expandafter\uppercase\expandafter{\romannumeral#1}}
\def\alphabetic#1{%
     \ifcase#1
          $\bullet$\err@badcountervalue{alphabetic}%
          \or a\or b\or c\or d\or e\or f\or g\or h\or i\or j\or k\or l\or m%
          \or n\or o\or p\or q\or r\or s\or t\or u\or v\or w\or x\or y\or z%
     \else$\bullet$\err@badcountervalue{alphabetic}%
     \fi}
\def\Alphabetic#1{\expandafter\uppercase\expandafter{\alphabetic{#1}}}
\def\symbols#1{%
     \ifcase#1
          $\bullet$\err@badcountervalue{symbols}%
          \or*\or\dag\or\ddag\or\S\or$\|$%
          \or**\or\dag\dag\or\ddag\ddag\or\S\S\or$\|\|$%
     \else$\bullet$\err@badcountervalue{symbols}%
     \fi}


\catcode`\^^?=13 \def^^?{\relax}

\def\trimleading#1\to#2{\edef#2{#1}%
     \expandafter\@trimleading\expandafter#2#2^^?^^?}
\def\@trimleading#1#2#3^^?{\ifx#2^^?\def#1{}\else\def#1{#2#3}\fi}

\def\trimtrailing#1\to#2{\edef#2{#1}%
     \expandafter\@trimtrailing\expandafter#2#2^^? ^^?\relax}
\def\@trimtrailing#1#2 ^^?#3{\ifx#3\relax\toks@={}%
     \else\def#1{#2}\toks@={\trimtrailing#1\to#1}\fi
     \the\toks@}

\def\trim#1\to#2{\trimleading#1\to#2\trimtrailing#2\to#2}

\catcode`\^^?=15


\long\def\additemL#1\to#2{\toks@={\^^\{#1}}\toks@ii=\expandafter{#2}%
     \xdef#2{\the\toks@\the\toks@ii}}

\long\def\additemR#1\to#2{\toks@={\^^\{#1}}\toks@ii=\expandafter{#2}%
     \xdef#2{\the\toks@ii\the\toks@}}

\def\getitemL#1\to#2{\expandafter\@getitemL#1\hack@#1#2}
\def\@getitemL\^^\#1#2\hack@#3#4{\def#4{#1}\def#3{#2}}

\message{font macros,}


\newdimen\rp@
\newcount\@@sizeindex \@@sizeindex=0
\newcount\@@factori
\newcount\@@factorii
\newcount\@@factoriii
\newcount\@@factoriv

\countdef\maxfam=18
\newfam\itfam
\newfam\bffam
\newfam\bfsfam
\newfam\bmitfam

\def\@mathfontinit{\count@=4
     \loop\textfont\count@=\nullfont
          \scriptfont\count@=\nullfont
          \scriptscriptfont\count@=\nullfont
          \ifnum\count@<\maxfam\advance\count@ by\@ne
     \repeat}

\def\@fontstyleinit{%
     \def\it{\err@fontnotavailable\it}%
     \def\bf{\err@fontnotavailable\bf}%
     \def\bfs{\err@bfstobf}%
     \def\bmit{\err@fontnotavailable\bmit}%
     \def\sc{\err@fontnotavailable\sc}%
     \def\sl{\err@sltoit}%
     \def\ss{\err@fontnotavailable\ss}%
     \def\tt{\err@fontnotavailable\tt}}

\def\@parameterinit#1{\rm\rp@=.1em \@getscaling{#1}%
     \let\^^\=\@doscaling\scalingskipslist
     \setbox\strutbox=\hbox{\vrule
          height.708\baselineskip depth.292\baselineskip width\z@}}

\def\@getfactor#1#2#3#4{\@@factori=#1 \@@factorii=#2
     \@@factoriii=#3 \@@factoriv=#4}

\def\@getscaling#1{\count@=#1 \advance\count@ by-\@@sizeindex\@@sizeindex=#1
     \ifnum\count@<0
          \let\@mulordiv=\divide
          \let\@divormul=\multiply
          \multiply\count@ by\m@ne
     \else\let\@mulordiv=\multiply
          \let\@divormul=\divide
     \fi
     \edef\@@scratcha{\ifcase\count@                {1}{1}{1}{1}\or
          {1}{7}{23}{3}\or     {2}{5}{3}{1}\or      {9}{89}{13}{1}\or
          {6}{25}{6}{1}\or     {8}{71}{14}{1}\or    {6}{25}{36}{5}\or
          {1}{7}{53}{4}\or     {12}{125}{108}{5}\or {3}{14}{53}{5}\or
          {6}{41}{17}{1}\or    {13}{31}{13}{2}\or   {9}{107}{71}{2}\or
          {11}{139}{124}{3}\or {1}{6}{43}{2}\or     {10}{107}{42}{1}\or
          {1}{5}{43}{2}\or     {5}{69}{65}{1}\or    {11}{97}{91}{2}\fi}%
     \expandafter\@getfactor\@@scratcha}

\def\@doscaling#1{\@mulordiv#1by\@@factori\@divormul#1by\@@factorii
     \@mulordiv#1by\@@factoriii\@divormul#1by\@@factoriv}


\newskip\headskip
\newskip\footskip

\def\typesize=#1pt{\count@=#1 \advance\count@ by-10
     \ifcase\count@
          \@setsizex\or\err@badtypesize\or
          \@setsizexii\or\err@badtypesize\or
          \@setsizexiv
     \else\err@badtypesize
     \fi}

\def\@setsizex{\getixpt
     \def\subsubscriptfonts{\vpt}%
          \def\subsubscriptsize{\vpt\@parameterinit{-8}}%
     \def\subscriptfonts{\viipt}\def\subscriptsize{\viipt\@parameterinit{-4}}%
     \def\footnotefonts{\viiipt}\def\footnotesize{\viiipt\@parameterinit{-2}}%
     \def\smallfonts{\ixpt}\def\smallsize{\ixpt\@parameterinit{-1}}%
     \def\normalfonts{\xpt}\def\normalsize{\xpt\@parameterinit{0}}%
     \def\bigfonts{\xiipt}\def\bigsize{\xiipt\@parameterinit{2}}%
     \def\Bigfonts{\xivpt}\def\Bigsize{\xivpt\@parameterinit{4}}%
     \def\biggfonts{\xviipt}\def\biggsize{\xviipt\@parameterinit{6}}%
     \def\Biggfonts{\xxipt}\def\Biggsize{\xxipt\@parameterinit{8}}%
     \def\tinyfonts{\vpt}\def\tinysize{\vpt\@parameterinit{-8}}%
     \def\HUGEFONTS{\xxvpt}\def\HUGESIZE{\xxvpt\@parameterinit{10}}%
     \normalsize\fixedskipslist}

\def\@setsizexii{\getxipt
     \def\subsubscriptfonts{\vipt}%
          \def\subsubscriptsize{\vipt\@parameterinit{-6}}%
     \def\subscriptfonts{\viiipt}%
          \def\subscriptsize{\viiipt\@parameterinit{-2}}%
     \def\footnotefonts{\xpt}\def\footnotesize{\xpt\@parameterinit{0}}%
     \def\smallfonts{\xipt}\def\smallsize{\xipt\@parameterinit{1}}%
     \def\normalfonts{\xiipt}\def\normalsize{\xiipt\@parameterinit{2}}%
     \def\bigfonts{\xivpt}\def\bigsize{\xivpt\@parameterinit{4}}%
     \def\Bigfonts{\xviipt}\def\Bigsize{\xviipt\@parameterinit{6}}%
     \def\biggfonts{\xxipt}\def\biggsize{\xxipt\@parameterinit{8}}%
     \def\Biggfonts{\xxvpt}\def\Biggsize{\xxvpt\@parameterinit{10}}%
     \def\tinyfonts{\vpt}\def\tinysize{\vpt\@parameterinit{-8}}%
     \def\HUGEFONTS{\xxvpt}\def\HUGESIZE{\xxvpt\@parameterinit{10}}%
     \normalsize\fixedskipslist}

\def\@setsizexiv{\getxiiipt
     \def\subsubscriptfonts{\viipt}%
          \def\subsubscriptsize{\viipt\@parameterinit{-4}}%
     \def\subscriptfonts{\xpt}\def\subscriptsize{\xpt\@parameterinit{0}}%
     \def\footnotefonts{\xiipt}\def\footnotesize{\xiipt\@parameterinit{2}}%
     \def\smallfonts{\xiiipt}\def\smallsize{\xiiipt\@parameterinit{3}}%
     \def\normalfonts{\xivpt}\def\normalsize{\xivpt\@parameterinit{4}}%
     \def\bigfonts{\xviipt}\def\bigsize{\xviipt\@parameterinit{6}}%
     \def\Bigfonts{\xxipt}\def\Bigsize{\xxipt\@parameterinit{8}}%
     \def\biggfonts{\xxvpt}\def\biggsize{\xxvpt\@parameterinit{10}}%
     \def\Biggfonts{\err@sizetoolarge\Biggfonts\HUGEFONTS}%
          \def\Biggsize{\err@sizetoolarge\Biggsize\HUGESIZE}%
     \def\tinyfonts{\vpt}\def\tinysize{\vpt\@parameterinit{-8}}%
     \def\HUGEFONTS{\xxvpt}\def\HUGESIZE{\xxvpt\@parameterinit{10}}%
     \normalsize\fixedskipslist}

\def\subsubscriptfonts{\vpt} \def\subsubscriptsize{\vpt\@parameterinit{-8}}
\def\subscriptfonts{\viipt}  \def\subscriptsize{\viipt\@parameterinit{-4}}
\def\footnotefonts{\viiipt}  \def\footnotesize{\viiipt\@parameterinit{-2}}
\def\smallfonts{\err@sizenotavailable\smallfonts}
                             \def\smallsize{\ixpt\@parameterinit{-1}}
\def\normalfonts{\xpt}       \def\normalsize{\xpt\@parameterinit{0}}
\def\bigfonts{\xiipt}        \def\bigsize{\xiipt\@parameterinit{2}}
\def\Bigfonts{\xivpt}        \def\Bigsize{\xivpt\@parameterinit{4}}
\def\biggfonts{\xviipt}      \def\biggsize{\xviipt\@parameterinit{6}}
\def\Biggfonts{\xxipt}       \def\Biggsize{\xxipt\@parameterinit{8}}
\def\tinyfonts{\vpt}         \def\tinysize{\vpt\@parameterinit{-8}}
\def\HUGEFONTS{\xxvpt}       \def\HUGESIZE{\xxvpt\@parameterinit{10}}

\message{document layout,}


\newtoks\everyoutput \everyoutput={}
\newdimen\depthofpage
\newcount\pagenum \pagenum=0

\newdimen\oddtopmargin  \newdimen\eventopmargin
\newdimen\oddleftmargin \newdimen\evenleftmargin
\newtoks\oddhead        \newtoks\evenhead
\newtoks\oddfoot        \newtoks\evenfoot

\def\topmargin{\afterassignment\@seteventop\oddtopmargin}
\def\leftmargin{\afterassignment\@setevenleft\oddleftmargin}
\def\head{\afterassignment\@setevenhead\oddhead}
\def\foot{\afterassignment\@setevenfoot\oddfoot}

\def\@seteventop{\eventopmargin=\oddtopmargin}
\def\@setevenleft{\evenleftmargin=\oddleftmargin}
\def\@setevenhead{\evenhead=\oddhead}
\def\@setevenfoot{\evenfoot=\oddfoot}

\def\pagenumstyle#1{\@setnumstyle\pagenum{#1}}

\newif\ifdraft
\def\draft{\drafttrue\leftmargin=.5in \overfullrule=5pt }

\def\outputstyle#1{\global\expandafter\let\expandafter
          \@outputstyle\csname#1output\endcsname
     \usename{#1setup}}

\output={\@outputstyle}

\def\normaloutput{\the\everyoutput
     \global\advance\pagenum by\@ne
     \ifodd\pagenum
          \voffset=\oddtopmargin \hoffset=\oddleftmargin
     \else\voffset=\eventopmargin \hoffset=\evenleftmargin
     \fi
     \advance\voffset by-1in  \advance\hoffset by-1in
     \count0=\pagenum
     \expandafter\shipout\pagebox
     \ifnum\outputpenalty>-\@MM\else\dosupereject\fi}

\newdimen\fullhsize
\newbox\leftpage
\newcount\leftpagenum
\newcount\outputpagenum \outputpagenum=0
\let\leftorright=L

\def\twoupoutput{\the\everyoutput
     \global\advance\pagenum by\@ne
     \if L\leftorright
          \global\setbox\leftpage=\leftline{\pagebox}%
          \global\leftpagenum=\pagenum
          \global\let\leftorright=R%
     \else\global\advance\outputpagenum by\@ne
          \ifodd\outputpagenum
               \voffset=\oddtopmargin \hoffset=\oddleftmargin
          \else\voffset=\eventopmargin \hoffset=\evenleftmargin
          \fi
          \advance\voffset by-1in  \advance\hoffset by-1in
          \count0=\leftpagenum \count1=\pagenum
          \shipout\vbox{\hbox to\fullhsize
               {\box\leftpage\hfil\leftline{\pagebox}}}%
          \global\let\leftorright=L%
     \fi
     \ifnum\outputpenalty>-\@MM
     \else\dosupereject
          \if R\leftorright
               \globaldefs=\@ne\head={\hfil}\foot={\hfil}\globaldefs=\z@
               \null\newpage
          \fi
     \fi}

\def\pagebox{\vbox{\makeheadline\pagebody\makefootline}}

\def\makeheadline{%
     \vbox to\z@{\baselinestretch=\@m
          \vskip\topskip\vskip-.708\baselineskip\vskip-\headskip
          \line{\vbox to\ht\strutbox{}%
               \ifodd\pagenum\the\oddhead\else\the\evenhead\fi}%
          \vss}%
     \nointerlineskip}

\def\pagebody{\vbox to\vsize{%
     \boxmaxdepth\maxdepth
     \ifvoid\topins\else\unvbox\topins\fi
     \depthofpage=\dp255
     \unvbox255
     \ifraggedbottom\kern-\depthofpage\vfil\fi
     \ifvoid\footins
     \else\vskip\skip\footins
          \footnoterule
          \unvbox\footins
          \vskip-\footnoteskip
     \fi}}

\def\makefootline{\baselineskip=\footskip
     \line{\ifodd\pagenum\the\oddfoot\else\the\evenfoot\fi}}


\newskip\abovechapterskip
\newskip\belowchapterskip
\newskip\abovesectionskip
\newskip\belowsectionskip
\newskip\abovesubsectionskip
\newskip\belowsubsectionskip

\def\chapterstyle#1{\global\expandafter\let\expandafter\@chapterstyle
     \csname#1text\endcsname}
\def\sectionstyle#1{\global\expandafter\let\expandafter\@sectionstyle
     \csname#1text\endcsname}
\def\subsectionstyle#1{\global\expandafter\let\expandafter\@subsectionstyle
     \csname#1text\endcsname}

\def\chapter#1{%
     \ifdim\lastskip=17sp \else\chapterbreak\vskip\abovechapterskip\fi
     \@chapterstyle{\ifblank\chapternumstyle\then
          \else\newchapternum=\next\chapternumformat\ \fi#1}%
     \nobreak\vskip\belowchapterskip\vskip17sp }

\def\section#1{%
     \ifdim\lastskip=17sp \else\sectionbreak\vskip\abovesectionskip\fi
     \@sectionstyle{\ifblank\sectionnumstyle\then
          \else\newsectionnum=\next\sectionnumformat\ \fi#1}%
     \nobreak\vskip\belowsectionskip\vskip17sp }

\def\subsection#1{%
     \ifdim\lastskip=17sp \else\subsectionbreak\vskip\abovesubsectionskip\fi
     \@subsectionstyle{\ifblank\subsectionnumstyle\then
          \else\newsubsectionnum=\next\subsectionnumformat\ \fi#1}%
     \nobreak\vskip\belowsubsectionskip\vskip17sp }


\let\TeXunderline=\underline
\let\TeXoverline=\overline
\def\underline#1{\relax\ifmmode\TeXunderline{#1}\else
     $\TeXunderline{\hbox{#1}}$\fi}
\def\overline#1{\relax\ifmmode\TeXoverline{#1}\else
     $\TeXoverline{\hbox{#1}}$\fi}

\def\baselinestretch{\afterassignment\@baselinestretch\count@}
\def\@baselinestretch{\baselineskip=\normalbaselineskip
     \divide\baselineskip by\@m\baselineskip=\count@\baselineskip
     \setbox\strutbox=\hbox{\vrule
          height.708\baselineskip depth.292\baselineskip width\z@}%
     \bigskipamount=\the\baselineskip
          plus.25\baselineskip minus.25\baselineskip
     \medskipamount=.5\baselineskip
          plus.125\baselineskip minus.125\baselineskip
     \smallskipamount=.25\baselineskip
          plus.0625\baselineskip minus.0625\baselineskip}

\def\\{\ifhmode\ifnum\lastpenalty=-\@M\else\hfil\penalty-\@M\fi\fi
     \ignorespaces}
\def\newpage{\vfil\break}

\def\lefttext#1{\par{\@text\leftskip=\z@\rightskip=\centering
     \noindent#1\par}}
\def\righttext#1{\par{\@text\leftskip=\centering\rightskip=\z@
     \noindent#1\par}}
\def\centertext#1{\par{\@text\leftskip=\centering\rightskip=\centering
     \noindent#1\par}}
\def\@text{\parindent=\z@ \parfillskip=\z@ \everypar={}%
     \spaceskip=.3333em \xspaceskip=.5em
     \def\\{\ifhmode\ifnum\lastpenalty=-\@M\else\penalty-\@M\fi\fi
          \ignorespaces}}

\def\beginleft{\par\@text\leftskip=\z@ \rightskip=\centering}
     
\def\beginright{\par\@text\leftskip=\centering\rightskip=\z@ }
     
\def\begincenter{\par\@text\leftskip=\centering\rightskip=\centering}

\def\beginnarrow{\defaultoption[\parindent]\@beginnarrow}
\def\@beginnarrow[#1]{\par\advance\leftskip by#1\advance\rightskip by#1}

\begingroup
\catcode`\[=1 \catcode`\{=11
\gdef\beginignore[\endgroup\bgroup
     \catcode`\e=0 \catcode`\\=12 \catcode`\{=11 \catcode`\f=12 \let\or=\relax
     \let\nd{ignor=\fi \let\}=\egroup
     \iffalse}
\endgroup

\long\def\marginnote#1{\leavevmode
     \edef\@marginsf{\spacefactor=\the\spacefactor\relax}%
     \ifdraft\strut\vadjust{%
          \hbox to\z@{\hskip\hsize\hskip.1in
               \vbox to\z@{\vskip-\dp\strutbox
                    \marginnoteformat
                    \vskip-\ht\strutbox
                    \noindent\strut#1\par
                    \vss}%
               \hss}}%
     \fi
     \@marginsf}


\newtoks\everybye \everybye={\par\vfil}
\outer\def\bye{\the\everybye
     \footnotecheck
     \prelabelcheck
     \streamcheck
     \supereject
     \TeXend}

\message{footnotes,}

\newcount\footnotenum \footnotenum=0
\newskip\footnoteskip
\let\@footnotelist=\empty

\def\footnotenumstyle#1{\@setnumstyle\footnotenum{#1}%
     \useafter\ifx{@footnotenumstyle}\symbols
          \global\let\@footup=\empty
     \else\global\let\@footup=\markup
     \fi}

\def\footnote{\footnotecheck\defaultoption[]\@footnote}
\def\@footnote[#1]{\@footnotemark[#1]\@footnotetext}

\def\footnotemark{\defaultoption[]\@footnotemark}
\def\@footnotemark[#1]{\let\@footsf=\empty
     \ifhmode\edef\@footsf{\spacefactor=\the\spacefactor\relax}\/\fi
     \ifnoarg#1\then
          \global\advance\footnotenum by\@ne
          \@footup{\footnotenumformat}%
          \edef\@@foota{\footnotenum=\the\footnotenum\relax}%
          \expandafter\additemR\expandafter\@footup\expandafter
               {\@@foota\footnotenumformat}\to\@footnotelist
          \global\let\@footnotelist=\@footnotelist
     \else\markup{#1}%
          \additemR\markup{#1}\to\@footnotelist
          \global\let\@footnotelist=\@footnotelist
     \fi
     \@footsf}

\def\footnotetext{%
     \ifx\@footnotelist\empty\err@extrafootnotetext\else\@footnotetext\fi}
\def\@footnotetext{%
     \getitemL\@footnotelist\to\@@foota
     \global\let\@footnotelist=\@footnotelist
     \insert\footins\bgroup
     \footnoteformat
     \splittopskip=\ht\strutbox\splitmaxdepth=\dp\strutbox
     \interlinepenalty=\interfootnotelinepenalty\floatingpenalty=\@MM
     \noindent\llap{\@@foota}\strut
     \bgroup\aftergroup\@footnoteend
     \let\@@scratcha=}
\def\@footnoteend{\strut\par\vskip\footnoteskip\egroup}

\def\footnoterule{\normalfonts
     \kern-.3em \hrule width2in height.04em \kern .26em }

\def\footnotecheck{%
     \ifx\@footnotelist\empty
     \else\err@extrafootnotemark
          \global\let\@footnotelist=\empty
     \fi}

\message{labels,}

\let\@@labeldef=\xdef
\newif\if@labelfile
\newwrite\@labelfile
\let\@prelabellist=\empty

\def\label#1#2{\trim#1\to\@@labarg\edef\@@labtext{#2}%
     \edef\@@labname{lab@\@@labarg}%
     \useafter\ifundefined\@@labname\then\else\@yeslab\fi
     \useafter\@@labeldef\@@labname{#2}%
     \ifstreaming
          \expandafter\toks@\expandafter\expandafter\expandafter
               {\csname\@@labname\endcsname}%
          \immediate\write\streamout{\noexpand\label{\@@labarg}{\the\toks@}}%
     \fi}
\def\@yeslab{%
     \useafter\ifundefined{if\@@labname}\then
          \err@labelredef\@@labarg
     \else\useif{if\@@labname}\then
               \err@labelredef\@@labarg
          \else\global\usename{\@@labname true}%
               \useafter\ifundefined{pre\@@labname}\then
               \else\useafter\ifx{pre\@@labname}\@@labtext
                    \else\err@badlabelmatch\@@labarg
                    \fi
               \fi
               \if@labelfile
               \else\global\@labelfiletrue
                    \immediate\write\sixt@@n{--> Creating file \jobname.lab}%
                    \immediate\openout\@labelfile=\jobname.lab
               \fi
               \immediate\write\@labelfile
                    {\noexpand\prelabel{\@@labarg}{\@@labtext}}%
          \fi
     \fi}

\def\putlab#1{\trim#1\to\@@labarg\edef\@@labname{lab@\@@labarg}%
     \useafter\ifundefined\@@labname\then\@nolab\else\usename\@@labname\fi}
\def\@nolab{%
     \useafter\ifundefined{pre\@@labname}\then
          \undefinedlabelformat
          \err@needlabel\@@labarg
          \useafter\xdef\@@labname{\undefinedlabelformat}%
     \else\usename{pre\@@labname}%
          \useafter\xdef\@@labname{\usename{pre\@@labname}}%
     \fi
     \useafter\newif{if\@@labname}%
     \expandafter\additemR\@@labarg\to\@prelabellist}

\def\prelabel#1{\useafter\gdef{prelab@#1}}

\def\ifundefinedlabel#1\then{%
     \expandafter\ifx\csname lab@#1\endcsname\relax}
\def\useiflab#1\then{\csname iflab@#1\endcsname}

\def\prelabelcheck{{%
     \def\^^\##1{\useiflab{##1}\then\else\err@undefinedlabel{##1}\fi}%
     \@prelabellist}}

\message{equation numbering,}

\newcount\chapternum
\newcount\sectionnum
\newcount\subsectionnum
\newcount\equationnum
\newcount\subequationnum
\newcount\figurenum
\newcount\subfigurenum
\newcount\tablenum
\newcount\subtablenum

\newif\if@subeqncount
\newif\if@subfigcount
\newif\if@subtblcount

\def\newchapternum{\newsectionnum=\z@\@resetnum\chapternum}
\def\newsectionnum{\newsubsectionnum=\z@\@resetnum\sectionnum}
\def\newsubsectionnum{\newequationnum=\z@\newfigurenum=\z@\newtablenum=\z@
     \@resetnum\subsectionnum}
\def\newequationnum{\newsubequationnum=\z@\@resetnum\equationnum}
\def\newsubequationnum{\@resetnum\subequationnum}
\def\newfigurenum{\newsubfigurenum=\z@\@resetnum\figurenum}
\def\newsubfigurenum{\@resetnum\subfigurenum}
\def\newtablenum{\newsubtablenum=\z@\@resetnum\tablenum}
\def\newsubtablenum{\@resetnum\subtablenum}

\def\@resetnum#1{\global\advance#1by1 \edef\next{\the#1\relax}\global#1}

\newchapternum=0

\def\chapternumstyle#1{\@setnumstyle\chapternum{#1}}
\def\sectionnumstyle#1{\@setnumstyle\sectionnum{#1}}
\def\subsectionnumstyle#1{\@setnumstyle\subsectionnum{#1}}
\def\equationnumstyle#1{\@setnumstyle\equationnum{#1}}
\def\subequationnumstyle#1{\@setnumstyle\subequationnum{#1}%
     \ifblank\subequationnumstyle\then\global\@subeqncountfalse\fi
     \ignorespaces}
\def\figurenumstyle#1{\@setnumstyle\figurenum{#1}}
\def\subfigurenumstyle#1{\@setnumstyle\subfigurenum{#1}%
     \ifblank\subfigurenumstyle\then\global\@subfigcountfalse\fi
     \ignorespaces}
\def\tablenumstyle#1{\@setnumstyle\tablenum{#1}}
\def\subtablenumstyle#1{\@setnumstyle\subtablenum{#1}%
     \ifblank\subtablenumstyle\then\global\@subtblcountfalse\fi
     \ignorespaces}

\def\eqnlabel#1{%
     \if@subeqncount
          \newsubequationnum=\next
     \else\newequationnum=\next
          \ifblank\subequationnumstyle\then
          \else\global\@subeqncounttrue
               \newsubequationnum=\@ne
          \fi
     \fi
     \label{#1}{\puteqnformat}(\puteqn{#1})%
     \ifdraft\rlap{\hskip.1in{\tt#1}}\fi}

\let\puteqn=\putlab

\def\equation#1#2{\useafter\gdef{eqn@#1}{#2\eqno\eqnlabel{#1}}}
\def\Equation#1{\useafter\gdef{eqn@#1}}

\def\putequation#1{\useafter\ifundefined{eqn@#1}\then
     \err@undefinedeqn{#1}\else\usename{eqn@#1}\fi}

\def\eqnseriesstyle#1{\gdef\@eqnseriesstyle{#1}}
\def\begineqnseries{\subequationnumstyle{\@eqnseriesstyle}%
     \defaultoption[]\@begineqnseries}
\def\@begineqnseries[#1]{\edef\@@eqnname{#1}}
\def\endeqnseries{\subequationnumstyle{blank}%
     \expandafter\ifnoarg\@@eqnname\then
     \else\label\@@eqnname{\puteqnformat}%
     \fi
     \aftergroup\ignorespaces}

\def\figlabel#1{%
     \if@subfigcount
          \newsubfigurenum=\next
     \else\newfigurenum=\next
          \ifblank\subfigurenumstyle\then
          \else\global\@subfigcounttrue
               \newsubfigurenum=\@ne
          \fi
     \fi
     \label{#1}{\putfigformat}\putfig{#1}%
     {\def\marginnoteformat{\tt}\marginnote{#1}}}

\let\putfig=\putlab

\def\figseriesstyle#1{\gdef\@figseriesstyle{#1}}
\def\beginfigseries{\subfigurenumstyle{\@figseriesstyle}%
     \defaultoption[]\@beginfigseries}
\def\@beginfigseries[#1]{\edef\@@figname{#1}}
\def\endfigseries{\subfigurenumstyle{blank}%
     \expandafter\ifnoarg\@@figname\then
     \else\label\@@figname{\putfigformat}%
     \fi
     \aftergroup\ignorespaces}

\def\tbllabel#1{%
     \if@subtblcount
          \newsubtablenum=\next
     \else\newtablenum=\next
          \ifblank\subtablenumstyle\then
          \else\global\@subtblcounttrue
               \newsubtablenum=\@ne
          \fi
     \fi
     \label{#1}{\puttblformat}\puttbl{#1}%
     {\def\marginnoteformat{\tt}\marginnote{#1}}}

\let\puttbl=\putlab

\def\tblseriesstyle#1{\gdef\@tblseriesstyle{#1}}
\def\begintblseries{\subtablenumstyle{\@tblseriesstyle}%
     \defaultoption[]\@begintblseries}
\def\@begintblseries[#1]{\edef\@@tblname{#1}}
\def\endtblseries{\subtablenumstyle{blank}%
     \expandafter\ifnoarg\@@tblname\then
     \else\label\@@tblname{\puttblformat}%
     \fi
     \aftergroup\ignorespaces}

\message{reference numbering,}

\newcount\referencenum \referencenum=0
\newcount\@@prerefcount \@@prerefcount=0
\newcount\@@thisref
\newcount\@@lastref
\newcount\@@loopref
\newcount\@@refseq
\newdimen\refnumindent
\let\@undefreflist=\empty

\def\referencenumstyle#1{\@setnumstyle\referencenum{#1}}

\def\referencestyle#1{\usename{@ref#1}}

\def\@refsequential{%
     \gdef\@refpredef##1{\global\advance\referencenum by\@ne
          \let\^^\=0\label{##1}{\^^\{\the\referencenum}}%
          \useafter\gdef{ref@\the\referencenum}{{##1}{\undefinedlabelformat}}}%
     \gdef\@reference##1##2{%
          \ifundefinedlabel##1\then
          \else\def\^^\####1{\global\@@thisref=####1\relax}\putlab{##1}%
               \useafter\gdef{ref@\the\@@thisref}{{##1}{##2}}%
          \fi}%
     \gdef\endputreferences{%
          \loop\ifnum\@@loopref<\referencenum
                    \advance\@@loopref by\@ne
                    \expandafter\expandafter\expandafter\@printreference
                         \csname ref@\the\@@loopref\endcsname
          \repeat
          \par}}

\def\@refpreordered{%
     \gdef\@refpredef##1{\global\advance\referencenum by\@ne
          \additemR##1\to\@undefreflist}%
     \gdef\@reference##1##2{%
          \ifundefinedlabel##1\then
          \else\global\advance\@@loopref by\@ne
               {\let\^^\=0\label{##1}{\^^\{\the\@@loopref}}}%
               \@printreference{##1}{##2}%
          \fi}
     \gdef\endputreferences{%
          \def\^^\####1{\useiflab{####1}\then
               \else\reference{####1}{\undefinedlabelformat}\fi}%
          \@undefreflist
          \par}}

\def\beginprereferences{\par
     \def\reference##1##2{\global\advance\referencenum by1\@ne
          \let\^^\=0\label{##1}{\^^\{\the\referencenum}}%
          \useafter\gdef{ref@\the\referencenum}{{##1}{##2}}}}
\def\endprereferences{\global\@@prerefcount=\the\referencenum\par}

\def\beginputreferences{\par
     \refnumindent=\z@\@@loopref=\z@
     \loop\ifnum\@@loopref<\referencenum
               \advance\@@loopref by\@ne
               \setbox\z@=\hbox{\referencenum=\@@loopref
                    \referencenumformat\enskip}%
               \ifdim\wd\z@>\refnumindent\refnumindent=\wd\z@\fi
     \repeat
     \putreferenceformat
     \@@loopref=\z@
     \loop\ifnum\@@loopref<\@@prerefcount
               \advance\@@loopref by\@ne
               \expandafter\expandafter\expandafter\@printreference
                    \csname ref@\the\@@loopref\endcsname
     \repeat
     \let\reference=\@reference}

\def\@printreference#1#2{\ifx#2\undefinedlabelformat\err@undefinedref{#1}\fi
     \noindent\ifdraft\rlap{\hskip\hsize\hskip.1in \tt#1}\fi
     \llap{\referencenum=\@@loopref\referencenumformat\enskip}#2\par}

\def\reference#1#2{{\par\refnumindent=\z@\putreferenceformat\noindent#2\par}}

\def\putref#1{\trim#1\to\@@refarg
     \expandafter\ifnoarg\@@refarg\then
          \toks@={\relax}%
     \else\@@lastref=-\@m\def\@@refsep{}\def\@more{\@nextref}%
          \toks@={\@nextref#1,,}%
     \fi\the\toks@}
\def\@nextref#1,{\trim#1\to\@@refarg
     \expandafter\ifnoarg\@@refarg\then
          \let\@more=\relax
     \else\ifundefinedlabel\@@refarg\then
               \expandafter\@refpredef\expandafter{\@@refarg}%
          \fi
          \def\^^\##1{\global\@@thisref=##1\relax}%
          \global\@@thisref=\m@ne
          \setbox\z@=\hbox{\putlab\@@refarg}%
     \fi
     \advance\@@lastref by\@ne
     \ifnum\@@lastref=\@@thisref\advance\@@refseq by\@ne\else\@@refseq=\@ne\fi
     \ifnum\@@lastref<\z@
     \else\ifnum\@@refseq<\thr@@
               \@@refsep\def\@@refsep{,}%
               \ifnum\@@lastref>\z@
                    \advance\@@lastref by\m@ne
                    {\referencenum=\@@lastref\putrefformat}%
               \else\undefinedlabelformat
               \fi
          \else\def\@@refsep{--}%
          \fi
     \fi
     \@@lastref=\@@thisref
     \@more}

\message{streaming,}

\newif\ifstreaming

\def\streamto{\defaultoption[\jobname]\@streamto}
\def\@streamto[#1]{\global\streamingtrue
     \immediate\write\sixt@@n{--> Streaming to #1.str}%
     \newwrite\streamout\immediate\openout\streamout=#1.str }

\def\streamfrom{\defaultoption[\jobname]\@streamfrom}
\def\@streamfrom[#1]{\newread\streamin\openin\streamin=#1.str
     \ifeof\streamin
          \expandafter\err@nostream\expandafter{#1.str}%
     \else\immediate\write\sixt@@n{--> Streaming from #1.str}%
          \let\@@labeldef=\gdef
          \ifstreaming
               \edef\@elc{\endlinechar=\the\endlinechar}%
               \endlinechar=\m@ne
               \loop\read\streamin to\@@scratcha
                    \ifeof\streamin
                         \streamingfalse
                    \else\toks@=\expandafter{\@@scratcha}%
                         \immediate\write\streamout{\the\toks@}%
                    \fi
                    \ifstreaming
               \repeat
               \@elc
               \input #1.str
               \streamingtrue
          \else\input #1.str
          \fi
          \let\@@labeldef=\xdef
     \fi}

\def\streamcheck{\ifstreaming
     \immediate\write\streamout{\pagenum=\the\pagenum}%
     \immediate\write\streamout{\footnotenum=\the\footnotenum}%
     \immediate\write\streamout{\referencenum=\the\referencenum}%
     \immediate\write\streamout{\chapternum=\the\chapternum}%
     \immediate\write\streamout{\sectionnum=\the\sectionnum}%
     \immediate\write\streamout{\subsectionnum=\the\subsectionnum}%
     \immediate\write\streamout{\equationnum=\the\equationnum}%
     \immediate\write\streamout{\subequationnum=\the\subequationnum}%
     \immediate\write\streamout{\figurenum=\the\figurenum}%
     \immediate\write\streamout{\subfigurenum=\the\subfigurenum}%
     \immediate\write\streamout{\tablenum=\the\tablenum}%
     \immediate\write\streamout{\subtablenum=\the\subtablenum}%
     \immediate\closeout\streamout
     \fi}


\def\err@badtypesize{%
     \errhelp={The limited availability of certain fonts requires^^J%
          that the base type size be 10pt, 12pt, or 14pt.^^J}%
     \errmessage{--> Illegal base type size}}

\def\err@badsizechange{\immediate\write\sixt@@n
     {--> Size change not allowed in math mode, ignored}}

\def\err@sizetoolarge#1{\immediate\write\sixt@@n
     {--> \noexpand#1 too big, substituting HUGE}}

\def\err@sizenotavailable#1{\immediate\write\sixt@@n
     {--> Size not available, \noexpand#1 ignored}}

\def\err@fontnotavailable#1{\immediate\write\sixt@@n
     {--> Font not available, \noexpand#1 ignored}}

\def\err@sltoit{\immediate\write\sixt@@n
     {--> Style \noexpand\sl not available, substituting \noexpand\it}%
     \it}

\def\err@bfstobf{\immediate\write\sixt@@n
     {--> Style \noexpand\bfs not available, substituting \noexpand\bf}%
     \bf}

\def\err@badgroup#1#2{%
     \errhelp={The block you have just tried to close was not the one^^J%
          most recently opened.^^J}%
     \errmessage{--> \noexpand\end{#1} doesn't match \noexpand\begin{#2}}}

\def\err@badcountervalue#1{\immediate\write\sixt@@n
     {--> Counter (#1) out of bounds}}

\def\err@extrafootnotemark{\immediate\write\sixt@@n
     {--> \noexpand\footnotemark command
          has no corresponding \noexpand\footnotetext}}

\def\err@extrafootnotetext{%
     \errhelp{You have given a \noexpand\footnotetext command without first
          specifying^^Ja \noexpand\footnotemark.^^J}%
     \errmessage{--> \noexpand\footnotetext command has no corresponding
          \noexpand\footnotemark}}

\def\err@labelredef#1{\immediate\write\sixt@@n
     {--> Label "#1" redefined}}

\def\err@badlabelmatch#1{\immediate\write\sixt@@n
     {--> Definition of label "#1" doesn't match value in \jobname.lab}}

\def\err@needlabel#1{\immediate\write\sixt@@n
     {--> Label "#1" cited before its definition}}

\def\err@undefinedlabel#1{\immediate\write\sixt@@n
     {--> Label "#1" cited but never defined}}

\def\err@undefinedeqn#1{\immediate\write\sixt@@n
     {--> Equation "#1" not defined}}

\def\err@undefinedref#1{\immediate\write\sixt@@n
     {--> Reference "#1" not defined}}

\def\err@nostream#1{%
     \errhelp={You have tried to input a stream file that doesn't exist.^^J}%
     \errmessage{--> Stream file #1 not found}}

\message{jyTeX initialization}

\everyjob{\immediate\write16{--> jyTeX version \fmtversion}%
     \edef\@@jobname{\jobname}%
     \edef\jobname{\@@jobname}%
     \settime
     \openin0=\jobname.lab
     \ifeof0
     \else\closein0
          \immediate\write16{--> Getting labels from file \jobname.lab}%
          \input\jobname.lab
     \fi}


\def\fixedskipslist{%
     \^^\{\topskip}%
     \^^\{\splittopskip}%
     \^^\{\maxdepth}%
     \^^\{\skip\topins}%
     \^^\{\skip\footins}%
     \^^\{\headskip}%
     \^^\{\footskip}}

\def\scalingskipslist{%
     \^^\{\p@renwd}%
     \^^\{\delimitershortfall}%
     \^^\{\nulldelimiterspace}%
     \^^\{\scriptspace}%
     \^^\{\jot}%
     \^^\{\normalbaselineskip}%
     \^^\{\normallineskip}%
     \^^\{\normallineskiplimit}%
     \^^\{\baselineskip}%
     \^^\{\lineskip}%
     \^^\{\lineskiplimit}%
     \^^\{\bigskipamount}%
     \^^\{\medskipamount}%
     \^^\{\smallskipamount}%
     \^^\{\parskip}%
     \^^\{\parindent}%
     \^^\{\abovedisplayskip}%
     \^^\{\belowdisplayskip}%
     \^^\{\abovedisplayshortskip}%
     \^^\{\belowdisplayshortskip}%
     \^^\{\abovechapterskip}%
     \^^\{\belowchapterskip}%
     \^^\{\abovesectionskip}%
     \^^\{\belowsectionskip}%
     \^^\{\abovesubsectionskip}%
     \^^\{\belowsubsectionskip}}


\def\twoupsetup{
     \topmargin=.75in
     \leftmargin=.5in
     \vsize=6.9in
     \hsize=4.75in
     \fullhsize=10in
     \let\draft=\relax}

\outputstyle{normal}                             

\def\marginnoteformat{\subscriptsize             
     \hsize=1in \baselinestretch=1000 \everypar={}%
     \tolerance=5000 \hbadness=5000 \parskip=0pt \parindent=0pt
     \leftskip=0pt \rightskip=0pt \raggedright}

\head={\ifdraft\normalfonts\it\hfil DRAFT\hfil   
     \llap{\number\day\ \monthword\month\ \militarytime}\else\hfil\fi}
\foot={\hfil\normalfonts\numstyle\pagenum\hfil}  

\normalbaselineskip=12pt                         
\normallineskip=0pt                              
\normallineskiplimit=0pt                         
\normalbaselines                                 

\topskip=.85\baselineskip
\splittopskip=\topskip
\headskip=2\baselineskip
\footskip=\headskip

\pagenumstyle{arabic}                            

\parskip=0pt                                     
\parindent=20pt                                  

\baselinestretch=1000                            


\chapterstyle{left}                              
\chapternumstyle{blank}                          
\def\chapterbreak{\newpage}                      
\abovechapterskip=0pt                            
\belowchapterskip=1.5\baselineskip               
     plus.38\baselineskip minus.38\baselineskip
\def\chapternumformat{\numstyle\chapternum.}     

\sectionstyle{left}                              
\sectionnumstyle{blank}                          
\def\sectionbreak{\vskip0pt plus4\baselineskip\penalty-100
     \vskip0pt plus-4\baselineskip}              
\abovesectionskip=1.5\baselineskip               
     plus.38\baselineskip minus.38\baselineskip
\belowsectionskip=\the\baselineskip              
     plus.25\baselineskip minus.25\baselineskip
\def\sectionnumformat{
     \ifblank\chapternumstyle\then\else\numstyle\chapternum.\fi
     \numstyle\sectionnum.}

\subsectionstyle{left}                           
\subsectionnumstyle{blank}                       
\def\subsectionbreak{\vskip0pt plus4\baselineskip\penalty-100
     \vskip0pt plus-4\baselineskip}              
\abovesubsectionskip=\the\baselineskip           
     plus.25\baselineskip minus.25\baselineskip
\belowsubsectionskip=.75\baselineskip            
     plus.19\baselineskip minus.19\baselineskip
\def\subsectionnumformat{
     \ifblank\chapternumstyle\then\else\numstyle\chapternum.\fi
     \ifblank\sectionnumstyle\then\else\numstyle\sectionnum.\fi
     \numstyle\subsectionnum.}


\footnotenumstyle{symbol}                       
\footnoteskip=0pt                                
\def\footnotenumformat{\numstyle\footnotenum}    
\def\footnoteformat{\footnotesize                
     \everypar={}\parskip=0pt \parfillskip=0pt plus1fil
     \leftskip=1em \rightskip=0pt
     \spaceskip=0pt \xspaceskip=0pt
     \def\\{\ifhmode\ifnum\lastpenalty=-10000
          \else\hfil\penalty-10000 \fi\fi\ignorespaces}}


\def\undefinedlabelformat{$\bullet$}             


\equationnumstyle{arabic}                        
\subequationnumstyle{blank}                      
\figurenumstyle{arabic}                          
\subfigurenumstyle{blank}                        
\tablenumstyle{arabic}                           
\subtablenumstyle{blank}                         

\eqnseriesstyle{alphabetic}                      
\figseriesstyle{alphabetic}                      
\tblseriesstyle{alphabetic}                      

\def\puteqnformat{\hbox{
     \ifblank\chapternumstyle\then\else\numstyle\chapternum.\fi
     \ifblank\sectionnumstyle\then\else\numstyle\sectionnum.\fi
     \ifblank\subsectionnumstyle\then\else\numstyle\subsectionnum.\fi
     \numstyle\equationnum
     \numstyle\subequationnum}}
\def\putfigformat{\hbox{
     \ifblank\chapternumstyle\then\else\numstyle\chapternum.\fi
     \ifblank\sectionnumstyle\then\else\numstyle\sectionnum.\fi
     \ifblank\subsectionnumstyle\then\else\numstyle\subsectionnum.\fi
     \numstyle\figurenum
     \numstyle\subfigurenum}}
\def\puttblformat{\hbox{
     \ifblank\chapternumstyle\then\else\numstyle\chapternum.\fi
     \ifblank\sectionnumstyle\then\else\numstyle\sectionnum.\fi
     \ifblank\subsectionnumstyle\then\else\numstyle\subsectionnum.\fi
     \numstyle\tablenum
     \numstyle\subtablenum}}


\referencestyle{sequential}                      
\referencenumstyle{arabic}                       
\def\putrefformat{\numstyle\referencenum}        
\def\referencenumformat{\numstyle\referencenum.} 
\def\putreferenceformat{
     \everypar={\hangindent=1em \hangafter=1 }%
     \def\\{\hfil\break\null\hskip-1em \ignorespaces}%
     \leftskip=\refnumindent\parindent=0pt \interlinepenalty=1000 }


\normalsize


\def\fmtversion{2.6M (June 1992)}

\catcode`\@=12

\input epsf
\footnotenumstyle{arabic}
\typesize=10pt
\magnification=1200
\baselineskip=17truept
\hsize=6truein\vsize=8.5truein
\sectionnumstyle{blank}
\chapternumstyle{blank}
\chapternum=1
\sectionnum=1
\pagenum=0

\def\begintitle{\pagenumstyle{blank}\parindent=0pt\begin{narrow}[0.4in]}
\def\endtitle{\end{narrow}\newpage\pagenumstyle{arabic}}


\def\beginexercise{\vskip 20truept\parindent=0pt\begin{narrow}[10
truept]}
\def\endexercise{\vskip 10truept\end{narrow}}


\def\eql#1{\eqno\eqnlabel{#1}}
\def\ref{\reference}
\def\peq{\puteqn}
\def\pref{\putref}

\def\mgn{\marginnote}
\def\bex{\begin{exercise}}
\def\eex{\end{exercise}}

\font\open=msbm10 
\font\goth=eufm10  

\def\mbox#1{{\leavevmode\hbox{#1}}}
\def\hspace#1{{\phantom{\mbox#1}}}
\def\oR{\mbox{\open\char82}}

\def\oZ{\mbox{\open\char90}}

\def\gf{\mbox{{\goth\char102}}}

\def\rS{{\rm S}}

\def\al{\alpha}
\def\be{\beta}

\def\Ga{\Gamma}

\def\om{\omega}

\def\si{\sigma}

\def\th{\theta}

\def\ze{\zeta}

\def\De{\Delta}

\def\Det{{\rm Det\,}}

\def\slice{{\rm slice}}

\def\zf{$\zeta$--function}


\def\frac#1/#2{\leavevmode\kern.1em
\raise.5ex\hbox{\the\scriptfont0 #1}\kern-.1em/\kern-.15em
\lower.25ex\hbox{\the\scriptfont0 #2}}
\def\sfrac#1/#2{\leavevmode\kern.1em
\raise.5ex\hbox{\the\scriptscriptfont0 #1}\kern-.1em/\kern-.15em
\lower.25ex\hbox{\the\scriptscriptfont0 #2}}

\def\gtorder{\mathrel{\raise.3ex\hbox{$>$}\mkern-14mu
             \lower0.6ex\hbox{$\sim$}}}
\def\ltorder{\mathrel{\raise.3ex\hbox{$<$}|mkern-14mu
             \lower0.6ex\hbox{\sim$}}}

\def\semidirprod{\rlap{\ss C}\raise1pt\hbox{$\mkern.75mu\times$}}
\def\for{\lower6pt\hbox{$\Big|$}}
\def\fish{\kern-.25em{\phantom{abcde}\over \phantom{abcde}}\kern-.25em}


\def\boxit#1{\vbox{\hrule\hbox{\vrule\kern3pt
        \vbox{\kern3pt#1\kern3pt}\kern3pt\vrule}\hrule}}
\def\dalemb#1#2{{\vbox{\hrule height .#2pt
        \hbox{\vrule width.#2pt height#1pt \kern#1pt
                \vrule width.#2pt}
        \hrule height.#2pt}}}

\def\ol{\overline}


\def\noin{\noindent}


\def\cosech{{\rm cosech\,}}
\def\etc{{\it etc. }}

\def\eg{{\it e.g. }}
\def\ie{{\it i.e. }}


  %

\def\3j#1#2#3#4#5#6{\left\lgroup\matrix{#1&#2&#3\cr#4&#5&#6\cr}
\right\rgroup}

\def\man{{\cal M}}

\def\caA{{\cal A}}
\def\caZ{{\cal Z}}
\def\caF{{\cal F}}

\def\caB{{\cal B}}

\def\m?{\mgn{?}}


\def\aop#1#2#3{{\it Ann. Phys.} {\bf {#1}} (19{#2}) #3}

\def\cmp#1#2#3{{\it Comm. Math. Phys.} {\bf {#1}} (19{#2}) #3}
\def\cqg#1#2#3{{\it Class. Quant. Grav.} {\bf {#1}} (19{#2}) #3}

\def\jmp#1#2#3{{\it J. Math. Phys.} {\bf {#1}} (19{#2}) #3}
\def\jpa#1#2#3{{\it J. Phys.} {\bf A{#1}} (19{#2}) #3}

\def\np#1#2#3{{\it Nucl. Phys.} {\bf B{#1}} (19{#2}) #3}
\def\pl#1#2#3{{\it Phys. Lett.} {\bf {#1}} (19{#2}) #3}

\def\pr#1#2#3{{\it Phys. Rev.} {\bf {#1}} (19{#2}) #3}
\def\prA#1#2#3{{\it Phys. Rev.} {\bf A{#1}} (19{#2}) #3}

\def\prD#1#2#3{{\it Phys. Rev.} {\bf D{#1}} (19{#2}) #3}

\def\prs#1#2#3{{\it Proc. Roy. Soc.} {\bf A{#1}} (19{#2}) #3}
\def\pcps#1#2#3{{\it Proc. Camb. Phil. Soc.} {\bf{#1}} (19{#2}) #3}

\def\dmj#1#2#3{{\it Duke Math. J.} {\bf {#1}} (19{#2}) #3}

\def\jdg#1#2#3{{\it J. Diff. Geom.} {\bf {#1}} (19{#2}) #3}
\def\jfa#1#2#3{{\it J. Func. Anal.} {\bf {#1}} (19{#2}) #3}

\def\jram#1#2#3{{\it J. f. reine u. Angew. Math.} {\bf {#1}} ({#2}) #3}
\def\ma#1#2#3{{\it Math. Ann.} {\bf {#1}} ({#2}) #3}

\def\pams#1#2#3{{\it Proc. Am. Math. Soc.} {\bf {#1}} (19{#2}) #3}

\def\tams#1#2#3{{\it Trans. Am. Math. Soc.} {\bf {#1}} (19{#2}) #3}

\begin{title}
\vglue 20truept
\vskip 100truept
\centertext {\Bigfonts \bf Casimir domains on a sphere}
\vskip 15truept
\centertext{J.S.Dowker\footnote{dowkeruk@yahoo.co.uk} }
\vskip 7truept
\centertext{\it Department of Theoretical Physics,\\
The University of Manchester, Manchester, England.}
\vskip 60truept
\centertext {Abstract}
\vskip 30truept
\begin{narrow}
A Casimir--type analysis of the effect of dividing the two--sphere by several  lines of latitude is done for conformally invariant Dirichlet and Neumann scalars and for spinors. An effective action combination is shown to have minima for symmetrical arrangements of the circles. For a domain of two caps and a slice, the Dirichlet slice expands to fill the whole sphere while the Neumann one adjusts to an angular separation of $57.92^\circ$.

The fermion expression is written in terms of the Weber class function, $\gf$, and a connection is noted with an earlier calculation of twisted scalar effective actions on the tetrahedron.

Illustrative special values are given.
\end{narrow}
\vskip 5truept
\vskip 75truept
\vfil
\end{title}
\pagenum=0
\section{\bf 1. Introduction}

The study of the effect of boundaries in quantum field theory became more active with the advent of the Casimir effect. Roughly, the question asked was -- given a manifold $\man$, (for simplicity take this as compact or with boundary), what is the effect of introducing an internal boundary that divides the space into two, or more, pieces, $\caA,\,\caB\ldots$? I write this schematically as $\man=\caA\cup\caB\cup\ldots$. The field would have boundary conditions applied at the inner boundaries (as well as at the outer one, which is taken as fixed). A natural quantity to consider would be, say, the free energy difference \footnote{ As in [\pref{DandK}] .},
$$
\De F(\man)\equiv F(\man)-F(\caA)-F(\caB)-\ldots\,.
\eql{casren}
$$
 One might then be interested in the consequences of moving or perturbing the introduced boundaries.
  
A considerable body of work has appeared over the years on various aspects of this question and I make no attempt to survey any of it. \footnote{ The analysis of boundary effects and obstacles is of ancient vintage in classical physics, acoustics, hydrodynamics and optics being prime examples. Also, boundary effects have long been a major interest in statistical physics.}

The same construct has appeared recently, [{\pref{PandW}], in a conformal, two--dimensional statistical/ probabilistic setting where $\man$ typically is a planar domain divided into parts by several, non-intersecting chords perpendicularly incident on the domain boundary.

Another analysis, [\pref{BKM}], deals with a closed Dirichlet curve on the conformal two--sphere and shows that, according to the logdet functional $F(\caA)+F(\caB)$, `critical' curves are circles on the metric sphere.  This general conclusion suggests the following more mundane programme.

The system I have in mind is an extension of earlier calculations in which conformal transformations played an important role. It bears a resemblance to the notion of the Casimir piston. Specifically, the domain $\man$ is the metric two--sphere and the domains, $\caA$, $\caB$ \etc\ are the caps and slices cut out by a series of parallel planes. I will deal mostly with the case of one slice and two polar caps. The answer depends on two angles which is easier to describe. The reason for the calculation is that it is exactly soluble, for free fields, and not very difficult amounting to algebraic combination of known quantities.

In the Casimir piston, one is dealing with the energy on a divided spatial section of space--time, see \eg [\pref{FKM}] for a recent treatment. Here I am looking at the one--loop effective action on $\man$ and have nothing to say about any physical relevance. All the ingredients are well known, somewhat elementary quantities in statistical physics and string theory but I am not aware of them being put together exactly in the way here.

\section{\bf 2. Basic equations.}
First I give some standard things. The functional determinant of the positive elliptic operator $D$ is defined
in terms of the \zf\ of $D$ by the usual relation,
$$
\log\Det D=-\ze'(0).
$$

For a conformally invariant field theory, the conformal behaviour
under infinitesimal Weyl rescalings, $g\to\bar g=\exp(-2\om)g$, is
controlled by the conformal anomaly, $\ze(0)$, up to zero modes. Integrated along a
conformal family of metrics, this anomaly yields the finite change in the
functional determinant (or, equivalently, the effective action
$W={1\over2}\log\Det D$),
$${1\over2}\log{\Det\overline D\over\Det D}=W[\bar g,g].
\eql{cchange}$$
In two dimensions, the cocycle function $W[\bar g, g]$ has been given by
L\"uscher, Symanzik and Weiss [\pref{Luscher}], Polyakov [\pref{Polyakov}]
when the boundary is empty and by Alvarez [\pref{Alvarez}] when not.

I shall be concerned with Dirichlet (D) and with Neumann (N) boundary conditions. The fields are conventional scalar and spinor ones and, although they are conformally covariant, I employ no CFT notions.

\section{\bf 3. Conformal transformations.}
The conformal transformations relevant here are those between
the sphere, $\rS^{2}$, Euclidean space, $\oR^{2}(\sim\oR^+
\times\rS^1)$ and the cylinder, $\oR\times\rS^1$.

These involve the equatorial stereographic projection, $\rS^{2}\to\oR^{2}$, expressed by giving the sphere metric in the conformally-flat form,
$$
d\si^2_{2}={4\over(1+r^2)^2}d{\bf r}^2.
\eql{stproj}$$
which, for example, connects the unit disk and the hemisphere.

An inverse stereographic projection takes an annulus to a {\it slice} of the sphere, $\rS^{2}$.

The metric on the cylinder is
$$
ds^2=d\tau^2+d\th^2=e^{-2\tau}\big(dr^2+r^2d\th^2\big)\,
\eql{cylproj}$$
with $r=\exp\tau$.

Because of the conformal invariance, the situation I consider could be described in any one of these manifolds. I choose the two--sphere because it treats zero and infinity in the same way and is more appealing.

\section{\bf 4. Cap and disk. One circle}

It is possible to rescale the unit disk and project it back onto the sphere thereby giving a spherical cap, as in [\pref{Dow2}]. The angle of the cap, $\th$, and the radius of the disk, $r$, are related by $\th=2\tan^{-1}r$.

The scalar cocycle functions are,
$$
W^N[\bar g,g]={1\over6}(1-\cos\th)-{1\over2}\log(1+\cos\th)-{1\over3}\log2
\eql{24}$$
and
$$
W^D[\bar g,g]={1\over6}\log2-{1\over3}(1-\cos\th).
\eql{14}$$
These enable the relation between cap and hemisphere,  \footnote{ The expression given in [\pref{Dow2}] is wrong and was corrected in [\pref{ApandD2}].} to be found,
$$
W^D_{\rm cap}(\th)-W^D_{\rm hemisphere}
=-{1\over3}\cos\th-{1\over6}\log\tan(\th/2)
\eql{15}$$
with, as determined by Weisberger,
$$
W^D_{\rm hemisphere}=-\ze_R'(-1)+{1\over8}-2\ze_R'(0)\,.
$$

The Neumann expressions are,
$$
W^N_{\rm cap}(\th)-W^N_{\rm hemisphere}={1\over6}\cos\th+
{1\over12}\log(1+\cos\th)+{5\over12}\log(1-\cos\th),
\eql{16}$$ with
$$
W^N_{\rm hemisphere}=-\ze_R'(-1)+{1\over8}+2\ze_R'(0).
$$

If there is just one plane (parallel to the equator) intersecting the sphere, \ie a total domain of just two supplementary caps, the symmetry behaviour of (\peq{15}) and (\peq{16}) show that for the Casimir renormalised quantity defined by,
  $$
 \De F(\th)\equiv W_{\rm sphere}-W_{\rm cap}(\pi-\th)-W_{\rm cap}(\th)\,,
  $$
one has
    $$\eqalign{
    &\De F^D(\th)= F^D_0\cr
    &\De F^N(\th)=F_0^N-\log\sin\th\,,
    }
    $$
 where $F_0$  are the universal constants,
  $$
      F_0^{N,D}= W_{\rm sphere}-2W^{N,D}_{\rm hemisphere}=\pm(W^D_{\rm hemisphere}-W^N_{\rm hemisphere})=\mp{1\over2}\log2\pi\,.
  $$

The D and N cases have a quite different behaviour.  For Dirichlet conditions $F$ is constant while, for Neuman, it has a minimum at the equator, $\th=90^\circ$, of value equal and opposite to the Dirichlet one.

\section{\bf5. Annulus and slice. Two circles}

Two lines  of latitude on the sphere bound, between them, a {\it slice}.
Conformal transformation of the plane into the two--sphere gives the effective action on a slice in terms of that on an annulus,
$$
W^D_{\slice}(\th_1,\th_2)=W^D_{\rm{annulus}}(r_1,r_2)
+{1\over3}\big(\cos \th_2-\cos\th_1\big)\,.
\eql{2slice}
$$
The colatitude of the slice runs from $\th_2$ to $\th_1$, ($\th_1>\th_2$) and $r_1$ and $r_2$ are the outer and inner radii of the annulus.

Furthermore, Weisberger, [\pref{Weisbergerb}], has obtained the effective action on the
annulus by conformal transformation from that on the cylinder, $I\times\rS^1$,
$$
W^D_{\rm annulus}(r_1,r_2)=W^D_{I\times\rS^1}-{1\over12}\log\big(r_1/r_2\big)\,.
$$

For pure Neumann conditions, one has the corresponding relation,
  $$\eqalign{
W^N_{\slice}&(\th_1,\th_2)\cr
&=W^N_{\rm{annulus}}(r_1,r_2)
+{1\over6}(\cos\th_1-\cos\th_2)+{1\over2}\log(1+\cos\th_1)(1+\cos\th_2)\,.
}
\eql{2slicen}
$$

I again note the geometrical relations,
$$\eqalign{
&r_1=\tan\th_1/2\,\,, r_2=\tan\th_2/2\cr
&L=\log R=\log{r_1\over r_2}>0.\cr
}
\eql{georel}$$
$L$ is the length of the cylinder \ie the length of the interval $I$ and $2L$ corresponds to an inverse temperature, $\beta$. Because it is suggestive, I will use $\be$ and refer to it as an inverse temperature even though there is nothing thermal about the arrangement.

All that remains in order to find the effective action  on the slice is a computation of that on the cylinder which is old and very familiar in statistical mechanics and string theory and elsewhere \footnote{ It is contained in the work of Epstein, [\pref{Epstein}], and Kronecker and that, earlier, of Greenhill and Hicks and, no doubt, elswhere.}. I give the answer in terms of a Lambert series,\footnote{ It is possible to go directly from the cylinder to the sphere, but I prefer to go via the plane.}
    $$
     W^D_{\rm slice}(\th_1,\th_2)=-\sum_{m=1}^\infty{1\over m(e^{m\be}-1)}+{1\over2}
     \log\big({\be\over2\pi}\big)-{\be\over12}+{1\over3}(\cos\th_2-\cos\th_1)
     \eql{slice}
    $$

The first term on the right corresponds to a statistical mechanical sum over states, with inverse temperature $\be=2\log R=2\log\big(\tan\th_1/2\,\cot\th_2/2\big)$. It is often expressed in terms of the Dedekind $\eta$ function, as I do later. The second term is a zero mode effect and the third one is $\be$ times the Casimir energy on the circle $E_0=-1/12$. The final term is a boundary geometric one.

The Casimir renormalised effective action, (\peq{casren}), can now be put together. The definition is,
  $$
      \De F^D(\th_1,\th_2)\equiv W_{\rm sphere}-W^D_{\rm cap}(\pi-\th_1)-W^D_{\rm cap}(\th_2)-W^D_{\rm slice}(\th_1,\th_2)\,.
      \eql{casren2}
  $$

As might be expected, there are cancellations between domains resulting in,
  $$
  \Xi^D(\be)\equiv \De F^D(\be)-F^D_0=\sum_{m=1}^\infty{1\over m(e^{m\be}-1)}-{1\over2}
     \log\big({\be\over2\pi}\big)
     \eql{deltaf}
  $$
which is, perhaps surprisingly, a function of only $\be$. \footnote{ The final term vanishes when $\be=2\pi$ giving a radii ratio $R=e^\pi$ which is Gelfond's constant.} I note that the terms proportional to $\be$ have disappeared. I will not combine the constant, $F^D_0$, with the expression on the right--hand--side.

The well--known inversion properties of the Lambert series summation \footnote{ These are those of the Dedekind $\eta$ function or the Epstein \zf\ or an Eisenstein series.} can be employed to rewrite (\peq{deltaf}) as,
  $$
  \Xi^D(\beta)=\sum_{m=1}^\infty{1\over m(e^{4\pi^2m/\be}-1)}+{1\over2}
     \log\big({\be\over2\pi}\big)-{\be\over24}+{\pi^2\over6\be}\,,
     \eql{deltaf2}
  $$
which is useful for small $\be$, ($\sim$ high temperatures) the final term being the Planck one.

It is notationally convenient to define the quantity,
   $$
      \caZ(\be)=e^{-\Xi(\be)}\,,
   $$
which can be looked upon as analogous to a partition function with a (`thermodynamic') free energy, $\Xi(\be)/\be$. \footnote{ I make the distinction because there is a tendency to refer to the logdet as a free energy. I prefer the term (one--loop) effective action for this.}

In terms of the Dedekind $\eta$--function, \footnote{ I write the argument of $\eta$ as $\beta$. More usually $q,=e^{-\be},$ is used},
   $$\eqalign{
      \caZ^D(\be) &= e^{\be/24}\bigg({\be\over2\pi}\bigg)^{1/2}\eta[\be]\cr
         &
         =e^{\pi^2/6\ol\be}\,\eta[\ol\be]\cr
         }
         \eql{dede}
   $$
where $\ol\beta$ is the  conjugate inverse temperature, $\beta\ol\beta=4\pi^2$.

The value $\be=\ol\be=2\pi$ is the symmetry point of the inversion when just the summation survives and equals one of Ramanujan's special values leading to,
   $$
     \caZ^D(2\pi)=e^{\pi/12}{\Ga(1/4)\over2\pi^{3/4}}\sim0.9981290699259585\sim1
   $$
related to elliptic functions (see the next section). The other special values lead to the following particular results
   $$\eqalign{
     &\caZ^D(\pi)={e^{\pi/24}\over2^{11/8}}{\Ga(1/4)\over\pi^{3/4}}\big)
     \sim 0.6752295518827370\sim1/\sqrt2\cr
     &\caZ^D(4\pi)={e^{\pi/6}\over2^{7/8}}{\Ga(1/4)\over\pi^{3/4}}
     \sim1.414208630509039\sim\sqrt2\cr
     &\caZ^D(8\pi)={e^{\pi/3}\over2^{13/16}}\big(\sqrt2-1\big)^{1/4}{\Ga(1/4)\over\pi^{3/4}}
     \sim1.999999999975676\sim2\,.\cr
     }
     \eql{spvalues}
   $$

The first two values are related by inversion, which, applied to the third, allows $\caZ^D(\pi/2)$ to be found. The trend of the values reflects the general limit (from the zero mode),
  $$
      \caZ^D(\be)\sim\bigg( {\be\over 2\pi}\bigg)^{1/2}\,, \quad \be\to\infty\,,
  $$
by a property of $\eta(\be)$.

To give an idea of the geometrical implications of the size of $\be$, consider a slice symmetrically placed about the equator. From (\peq{georel}), the angle difference between the rims, $2\al$, is related to $\be$ by $\al(\be)=\tan^{-1}e^{\be/4}$ giving $\al(\pi/2)\approx55.97^\circ$, $\al(\pi)\approx65.49^\circ$, $\al(2\pi)\approx78.25^\circ$, $\al(4\pi)\approx87.53^\circ$ and $\al(8\pi)\approx89.89^\circ$, the last almost the entire sphere.

The Neumann slice expressions are a little more involved because of the zero mode and, unlike the Dirichlet case, $\De F^N$ cannot be written as just a function of $\be$ although cancellations do occur when constructing the Casimir combination, (\peq{casren}). For brevity, I do not expose the formulae.

\section{\bf 6. Evaluation}

The values of the temperature in the previous section are special cases of a classic formula in terms of elliptic functions which reads, for the $\eta$--function, quite generally, \footnote{This formula is due to Jacobi.}
  $$
    \eta^{6}(\be)=2k\,{k'}\, { K^3[\be]\over\pi^3}
    \eql{ell}
  $$
where $K$ is the complete elliptic function of the first kind with modulus, $k$, and $k'$ is the conjugate modulus with $k^2+{k'}^2=1$. The connection with the inverse temperature is,
$$
    \be=2\pi{K'\over K}\,.
$$
$K$ is a function of $\be$ and is more usually written $K(k), =K[\be]$.   The interchange $k\leftrightarrow k'$ corresponds to inversion.

We are thus presented with a choice of numerical evaluations -- either perform the statistical sums in (\peq{deltaf}) and (\peq{deltaf2}) as they stand, use an existing CAS for $\eta$ or for the elliptic function in (\peq{ell}). There is no particular advantage in the latter except that for special, `singular' values of the modulus, $k=k_N$, the ratio, $K'/K$,  is the square root of a rational number, $N$, and  $K$ is known in terms of Gamma functions on the rationals, the examples in the previous section being typical (and relatively simple).

This route was pursued in [\pref{DandK1}] in connection with statistical mechanics on the spherical cylinder T$\,\times\,$S$^d$ (for odd $d$) and relevant references can be found there  as well as a few, selected singular moduli with their corresponding $K$ values. The resulting final expressions are, perhaps, more curious than practical.

Calculating at high special values of $\be$ enables one to obtain accurate approximations for Gelfond's constant, $e^\pi$, and $\Ga(1/4)$ (for example) in terms of algebraic numbers.

\section{\bf7. Equilibrium. One and two circles}

I {\it assume} now that the system of domains (caps and slices) seeks to minimise the effective action combination by adjusting the latitudes of the bounding circles. This is the only deformation that is easy to make.

The one D--circle has a uniform $\De F$ and so is free to rest anywhere, By contrast, the single N--circle will settle on the equator.

For more circles, it is to be expected that any global equilibrium state will be symmetrical about the equator and, indeed, calculation reveals that with two circles, for Dirichlet conditions one circle preferentially shrinks to the north pole and the other to the south, the effective action diverging negatively. Roughly speaking, the circles appear to repel each other and be attracted by the poles, although the behaviour is really a property of the entire system.

The Neumann situation is somewhat more interesting. The circles act as though repelled
both mutually and by the poles. The result is a finite global minimum with the two circles resting at supplementary colatitudes, $\approx 61.04^\circ$ and $\approx 118.96^\circ$ \ie approximately one radian angular separation.

\section{\bf 8. Multislice. More circles}

I alter the notation to write $\Xi^D(\be_{i,i+1})$  for the single slice quantity for angles $\th_{i}>\th_{i+1}$ and consider the situation when several planes parallel to the equator divide up the sphere. For $p$ number of planes, there will be $p-1$ slices and two caps. Combining the effective actions the intermediate boundary contributions are found to cancel for Dirichlet conditions leaving,
$$
 \caZ^D(\{\be_{i,i+1}\})=\prod_{i=1}^{p-1}\caZ^D(\be_{i,i+1})\,.
 \eql{multiD}
$$

I will describe the case of $p=3$, \ie two slices,
$$
 \caZ^D(\be_{1,2},\be_{2,3})=\caZ^D(\be_{1,2})\caZ^D(\be_{2,3})\,,
$$
 which is a function of three angles, $\th_1>\th_2>\th_3$. Again, numerical evaluation shows that the outer two circles collapse to the nearest poles while the centre one stabilises to the equator. This is the general behaviour for an odd number of circles. For an even number, the circles divide into two equal sets which collapse separately onto the poles.

 In slightly more detail, take the case of three circles. Fix the positions of the  top and bottom circles and allow the middle one to vary. It is found that the effective action has a minimum and, accordingly, the middle circle will settle nearer, it turns out, to whichever circle is closest to its pole. Now allow the top and bottom circles to vary, the middle one will follow the resulting moving minimum with the value of the effective action being lowest when the top and bottom circles are actually symmetrically placed about the equator. This lowest minimum becomes even lower as the top and bottom circles shrink to their poles.

 Again, the Neumann case is a little more complicated and cannot be written as neatly as (\peq{multiD}). Numerical analysis reveals, more interestingly, a finite global minimum with the middle circle at the equator ($\th_2=90^\circ$) and the other two symmetrically placed with $\th_1\approx136.14^\circ$ and $\th_3\approx43.86^\circ$.

 The numerical analysis can be carried as far as one wishes without difficulty. It shows that as more and more N--circles are introduced they are pushed, symmetrically, towards the poles becoming more and more squashed together. Since there is no obvious physical content in the set up, I do not pursue this any further.
\begin{ignore}
\epsfxsize=5truein \epsfbox{neufig2.eps}
\end{ignore}

\section{\bf9. Fermions}

The entire analysis can be repeated for a massless Dirac field which is conformally invariant. This requires the use of local `mixed' boundary conditions in which half the components satisfy Dirichet conditions and the rest Neumann (actually Robin). There are two such conditions but they lead to the same spectral data, basically as a consequence of the mixing.

The general procedure is exactly as in the scalar case and leads to,
   $$
   W^F_{\rm cap}(\th)-W^F_{\rm hemisphere}=-{1\over12}\cos\th
   -{1\over6}\log\tan\th/2\,,
   \eql{spcaphs}
   $$
   $$
    W^F_{\rm annulus}(r_1,r_2)-W^F_{I\times S^1}(L,a)=-{1\over12}\log r_1/r_2
    \eql{spanncyl}
   $$
 and
   $$
   W^F_{\rm slice}(\th_1,\th_2)-W^F_{\rm annulus}(r_1,r_2)={1\over12}(\cos\th_2-\cos\th_1)\,.
   \eql{spsliceann}
   $$

 The fermion cylinder effective action can be written in two ways, [\pref{DandA2}],
   $$\eqalign{
        W^F_{I\times S^!}(L/a)&=-{L\over12a}+\sum_{m=1}^\infty{(-1)^m\over m}\,\cosech mL/a\cr
        &=-{L\over12a}+\sum_{m=1}^\infty{1\over m}\big(\cosech 2mL/a-\cosech mL/a\big)\,.
        }
        \eql{Wcyl}
   $$
Because the conformal anomaly is zero, this depends only on the shape of the cylinder and not on its size. There is also no zero mode.

I note, before proceeding, the inversion symmetry due to Cauchy, \eg [\pref{Berndt2}],
   $$
       \sum_{m=1}^\infty{(-1)^m\over m}\big(\cosech \mu m-\cosech \nu m\big)=
       {1\over12}(\mu-\nu)\,,
       \eql{invs}
   $$
 with $\mu\nu=\pi^2$.
 The cylinder effective action has a maximum at the symmetry point of the inversion \ie when $L/a=\pi$. For the annulus the maximum is at
 $L/a\approx2.42569914$ equivalent to the ratio of radii, $r_1/r_2\approx11.3101$. The corresponding  scalar values were obtained some time ago by Osgood {\it et al}, [\pref{Osgood}].

As in the scalar case, elliptic quantities can be introduced, most easily by noting that, \eg\ [\pref{zucker2}],
    $$
       \sum_{m=1}^\infty{(-1)^m\over m}\cosech \mu m=-2\,\log Q_2(q)\,,
    $$
where $Q_2$ is one of Jacobi's products,
   $$
         Q_2(q)=\prod_{n=1}^\infty (1+q^{2n-1})\,,
   $$
and $q=e^{-\mu}$ is the nome. Here, $\mu=L/a$.  In terms of $Q_2$, the inversion symmetry (\peq{invs}) reads,
  $$
       q^{-1/24}\,Q_2(q)=q'^{-1/24}\,Q_2(q')\,.
  $$
 
 This invariant combination should be looked upon as the spinor analogue of the Dedekind eta function and so I can set,
    $$
     \gf(q)=q^{-1/24}\,Q_2(q)\,,
     \eql{spinded}
    $$
because this is seen to be one of Weber's class functions, [\pref{Weber}]. \footnote{ I remark that the inversion invariance of $\gf$ is really due to Cauchy. Ramanujan also gave this result.}

 This function has already appeared, with unrecognised history, in connection with QED effective actions in external fields, [\pref{KSX}]. It can be expressed neatly in terms of the Dedekind function by the ratio \footnote{ I write $\eta$ as a function of the nome. The more usual variable is $\tau=(i\log q)/\pi$.},
   $$
       \gf(q)={\eta^2(q)\over\eta(q^2)\eta(\sqrt q)}\,,
       \eql{spinded2}
   $$
obtained on rearrangement of the $q$ products.  \footnote{ There are  numerous ways of expressing spinor quantities in terms of scalar ones, (`bosonisation'),  \eg [\pref{DandK1}].}

I can now write the fermion cylinder effective action as,
   $$
     W_{I\times S^1}^F[q]=-2\log\gf(q)\,,\quad q=e^{-L/a}\,.
   $$

It will be no surprise that we have encountered quantities well known in CFT and string theory, \eg [\pref{IZ,DMS}],  with a notation similar to that in the mathematical [\pref{YiZ}].

The spin--structure \footnote{ See Seiberg and Witten [\pref{SandWi}].} under discussion here corresponds to the (N-S,N-S) string sector. The modes on the circle section of the cylinder are the natural ones that arise on treating S$^1$ as the $d=1$ $d$--sphere, S$^d$.
 
The other two Weber functions, $\gf_1$ and $\gf_2$, correspond to the (R,N-S) and (N-S,R) sectors and are interchanged on inversion. There exist interesting relations between $\gf$, $\gf_1$, $\gf_2$ and the other modular functions.
 
Incidentally, the three types of fermion effective action on the cylinder agree, up to the statistics sign, with those evaluated some time ago, [\pref{dowtetra}] App.B, for scalar fields on a tetrahedron and correspond to the three types of twisting around the vertices of the tetrahedron allowed by its topology. \footnote{ The tetrahedron can be regarded as a reflective $\oZ_2$ orbifolded torus and is sometimes referred to as a `pillowcase'.}
 
 Equation (\peq{spinded2}) is convenient for numerical computation, or the defining summation can be performed. Another, possibly useful expression is,
    $$
    \gf(q) =\sqrt{{\th_3(q)\over\eta(q)}}\,.
    $$

 Alternatively, the list of class--invariants in Weber, [\pref{Weber}], at special values of the argument, gives many `exact' values.  Typical examples, chosen at random, are,
   $$\eqalign{
     Q_2(e^{-\sqrt3\pi})&=2^{1/3}\,e^{-\sqrt3\pi/24}\approx 0.00432413965737\cr
     Q_2(e^{-5\pi})&={(1+\sqrt5)\over8^{1/4}}\,e^{-5\pi/24}\approx0.00008069626206\,,
     }
   $$
 but there is no particular advantage in these apart from amusement. I remark on the absence of Gamma functions.
 
 Equivalently one can employ existing lists of singular moduli after noting the relation,
   $$
      \log\gf(q)={1\over6}\log2-{1\over12}\log kk'\,,
   $$
(which neatly exhibits the invariance under inversion, $k\leftrightarrow k'$).   
   
 \section{\bf10. Fermionic domains on the sphere}

 The spinor Casimir renormalised effective action can now be assembled from the ingredients in the preceding section. The situation is similar to the Dirichlet scalar case. Considerable cancellations occur.

 For a single circle, because of the antisymmetry in (\peq{spcaphs}) $\De\caF$ is, as in the scalar case, independent of the position of the circle. For one slice the expression obtained is,
   $$
    \De \caF^F(\th_1,\th_2)=-{L\over12 a}+2\log\gf(e^{-L/a})\,,\quad L/a=\log(\tan\th_1\cot\th_2)\,,
   $$
 where I have used the fact that, because of the mixed mode structure, the Dirac \zf\ on the hemisphere is half that on the sphere. The vacuum energies and boundary terms have all cancelled.

 Minimising $\De\caF^F(\th_1,\th_2)$ leads to the same behaviour as in the scalar Dirichlet case with $\th_1\to\pi$ (the south pole) and $\th_2\to0$ (the north). The same conclusion holds for the multislice and needs no further comment.

\section{\bf11. Conclusion}

A straightforward analysis of the effect of dividing the two--sphere by circles of latitude is pursued for conformally invariant Dirichlet and Neumann scalars and spinors. The relevant effective action combination is shown to have minima for symmetrical arrangements of the circles. The expressions have been written in terms of elliptic function quantities and some special values given.

A simple extension that could be made is an arrangement of a D and an N cap with a D-N slice in between. This is not north--south symmetric.

There is no difficulty continuing the calculation to the spheres S$^3$, $S^4$ and, in principle, higher.

\vskip 10truept
\noin{\bf{References}}
\vskip 5truept
\begin{putreferences}
    \ref{BKM}{D.Burghelea, T.Kappeler and P.McDonald, {\it On the Functional logdet and Related Flows on the Space of Closed Embedded Curves on {\rm S}$^2$}, {\it J.Func.Anal} (1994) 440.}
     \ref{PandW}{E.Peltiola and Y.Wang {\it Large deviations of multichordal SLE0+, real rational functions and zeta-regularized determinants of Laplacians}, arXiv:2006.08574. }
     \ref{YiZ}{N.Yui and D.Zagier {\it Math. Comp} {\bf 66} (1997) 1645.}
     \ref{KSX}{H.Kleinert, E.Strobel and S-S.Xue {\it Phys. Rev.} D {\bf 88} (2013) 025049     .}
        \ref{DMS}{P.Di Francesco, P.Mathieu and D.S\'en\'echal {\it Conformal Field Theory}. Springer (New York) 1997.}
     \ref{Berndt2}{B.C.Berndt \jram{303/304}{1978}{332}.}
   \ref{SandWi}{N.Seiberg and E.Witten \np{276}{86}{272}.}
  \ref{dowtetra}{J.S.Dowker {\it Vacuum and magnetic field effects on a tetrahedron and other flat Riemann surfaces}, {\it Phys.Rev.} D{\bf{40}} (1989) {1938}.}
\ref{IZ}{C. Itzykson and J.B. Zuber,  \np{275}{86}{580}.}
  \ref{FKM}{G. Fucci, K. Kirsten and J.M. Mu\~noz-Caste\~neda {\it Casimir pistons with generalized boundary conditions: a step forward},  arXiv:1906.08486. }
\ref{DandK1}{J.S.Dowker,J.S. and K.Kirsten {\it Elliptic functions and
  temperature inversion symmetry on spheres}, \np {638} {2002} 405, arXiv:hep-th/0205029.}
  \ref{ZandR}{Zucker,I.J. and Robertson,M.M.
  {\it Math.Proc.Camb.Phil.Soc} {\bf 95 }(1984) 5.}
  \ref{JandZ1}{Joyce,G.S. and Zucker,I.J.
  {\it Math.Proc.Camb.Phil.Soc} {\bf 109 }(1991) 257.}
  \ref{JandZ2}{Zucker,I.J. and Joyce.G.S.
  {\it Math.Proc.Camb.Phil.Soc} {\bf 131 }(2001) 309.}
  \ref{zucker2}{I.J.Zucker {\it SIAM J.Math.Anal.} {\bf 10} (1979) 192.}
  \ref{BandZ}{Borwein,J.M. and Zucker,I.J. {\it IMA J.Math.Anal.} {\bf 12}
  (1992) 519.}
\ref{BandB}{Borwein,J.M. and Borwein,P.B. {\it Pi and the AGM}, Wiley,
  New York, 1998.}
\ref{DandA}{Dowker,J.S. and Apps, J.S. \cqg{15}{1998}{1121}.}
 \ref{ApandD}{Apps,J.S. and Dowker,J.S. \cqg{15}{1998}{1121}.}
 \ref{ApandD2}{J.S.Dowker and J.S.Apps \cqg{12}{95}{1363}.}
\ref{DandA2}{J.S.Dowker and J.S.Apps, {\it Functional determinants on
certain domains}. {\it Int. J. Mod. Phys. D} {\bf 5} (1996) 799; hep-th/9506204.}
\ref{Apps}{Apps,J.S. PhD thesis (University of Manchester, 1996).}
\ref{Weber}{H.Weber, {\it Lehrbuch der Algebra, Bd.III}, Vieweg (Braunschweig) (1908).}
\ref{KCD}{G.Kennedy, R.Critchley and J.S.Dowker \aop{125}{80}{346}.}
\ref{Donnelly}{H.Donnelly \ma{224}{1976}161.}
\ref{Fur2}{D.V.Fursaev {\sl Spectral geometry and one-loop divergences on
manifolds with conical singularities}, JINR preprint DSF-13/94,
hep-th/9405143.}
\ref{HandE}{S.W.Hawking and G.F.R.Ellis {\sl The large scale structure of
space-time} Cambridge University Press, 1973.}
\ref{DandK}{J.S.Dowker and G.Kennedy \jpa{11}{78}{895}.}
\ref{ChandD}{Peter Chang and J.S.Dowker \np{395}{93}{407}.}
\ref{FandM}{D.V.Fursaev and G.Miele \pr{D49}{94}{987}.}
\ref{Dowkerccs}{J.S.Dowker \cqg{4}{87}{L157}.}
\ref{BandH}{J.Br\"uning and E.Heintze \dmj{51}{84}{959}.}
\ref{Cheeger}{J.Cheeger \jdg{18}{83}{575}.}
\ref{SandW}{K.Stewartson and R.T.Waechter \pcps{69}{71}{353}.}
\ref{CandW}{H.S.Carslaw and J.C.Jaeger {\it The conduction of heat
in solids}
Oxford, The Clarendon Press 1959.}
\ref{BandH}{H.P.Baltes and E.M.Hilf {\it Spectra of finite systems}.}
\ref{Epstein}{P.Epstein \ma{56}{1903}{615}.}
\ref{Kennedy1}{G.Kennedy \pr{D23}{81}{2884}.}
\ref{Kennedy2}{G.Kennedy PhD thesis, Manchester (1978).}
\ref{Kennedy3}{G.Kennedy \jpa{11}{78}{L173}.}
\ref{Luscher}{M.L\"uscher, K.Symanzik and P.Weiss \np {173}{80}{365}.}
\ref{Polyakov}{A.M.Polyakov \pl {103}{81}{207}.}
\ref{Bukhb}{L.Bukhbinder, V.P.Gusynin and P.I.Fomin {\it Sov. J. Nucl.
 Phys.} {\bf 44} (1986) 534.}
\ref{Alvarez}{O.Alvarez \np {216}{83}{125}.}
\ref{DandS}{J.S.Dowker and J.P.Schofield \jmp{31}{90}{808}.}
\ref{Dow1}{J.S.Dowker \cmp{162}{94}{633}.}
\ref{Dow2}{J.S.Dowker \cqg{11}{94}{557}.}
\ref{Dow3}{J.S.Dowker \jmp{35}{94}{4989}; erratum {\it ibid}, Feb.1995.}
\ref{Dow5}{J.S.Dowker {\it Heat-kernels and polytopes} To be published}
\ref{Dow6}{J.S.Dowker \pr{D50}{94}{6369}.}
\ref{Dow7}{J.S.Dowker \pr{D39}{89}{1235}.}
\ref{BandG}{T.P.Branson and P.B.Gilkey\tams{344}{94}{479}.}
\ref{Schofield}{J.P.Schofield Ph.D.thesis, University of Manchester, (1991).}
\ref{Barnesa}{E.W.Barnes {\it Trans. Camb. Phil. Soc.} {\bf 19} (1903)
374.}
\ref{BandG2}{T.P.Branson and P.B.Gilkey {\it Comm. Partial Diff. Equations}
{\bf 15} (1990) 245.}
\ref{Pathria}{R.K.Pathria {\it Suppl.Nuovo Cim.} {\bf 4} (1966) 276.}
\ref{Baltes}{H.P.Baltes \prA{6}{72}{2252}.}
\ref{Spivak}{M.Spivak {\it Differential Geometry} vols III, IV, Publish
or Perish, Boston, 1975.}
\ref{Eisenhart}{L.P.Eisenhart {\it Differential Geometry}, Princeton
University Press, Princeton, 1926.}
\ref{Moss}{I.Moss \cqg{6}{89}{659}.}
\ref{Barv}{A.O.Barvinsky, Yu.A.Kamenshchik and I.P.Karmazin \aop {219}
{92}{201}.}
\ref{Kam}{Yu.A.Kamenshchik and I.V.Mishakov {\it Int. J. Mod. Phys.}
{\bf A7} (1992) 3265.}
\ref{Death}{P.D.D'eath and G.V.M.Esposito \prD{43}{91}{3234}.}
\ref{Rich}{K.Richardson \jfa{122}{94}{52}.}
\ref{Osgood}{B.Osgood, R.Phillips and P.Sarnak \jfa{80}{88}{148}.}
\ref{BCY}{T.P.Branson, S.-Y. A.Chang and P.C.Yang \cmp{149}{92}{241}.}
\ref{Vass}{D.V.Vassilevich.{\it Vector fields on a disk with mixed
boundary conditions} gr-qc /9404052.}
\ref{MandP}{I.Moss and S.Poletti \pl{B333}{94}{326}.}
\ref{Kam2}{G.Esposito, A.Y.Kamenshchik, I.V.Mishakov and G.Pollifrone
\prD{50}{94}{6329}.}
\ref{Aurell1}{E.Aurell and P.Salomonson \cmp{165}{94}{233}.}
\ref{Aurell2}{E.Aurell and P.Salomonson {\it Further results on functional
determinants of laplacians on simplicial complexes} hep-th/9405140.}
\ref{BandO}{T.P.Branson and B.\O rsted \pams{113}{91}{669}.}
\ref{Elizalde1}{E.Elizalde, \jmp{35}{94}{3308}.}
\ref{BandK}{M.Bordag and K.Kirsten {\it Heat-kernel coefficients of
the Laplace operator on the 3-dimensional ball} hep-th/9501064.}
\ref{Waechter}{R.T.Waechter \pcps{72}{72}{439}.}
\ref{GRV}{S.Guraswamy, S.G.Rajeev and P.Vitale {\it O(N) sigma-model as
a three dimensional conformal field theory}, Rochester preprint UR-1357.}
\ref{CandC}{A.Capelli and A.Costa \np {314}{89}{707}.}
\ref{IandZ}{C.Itzykson and J.-B.Zuber \np{275}{86}{580}.}
\ref{BandH}{M.V.Berry and C.J.Howls \prs {447}{94}{527}.}
\ref{DandW}{A.Dettki and A.Wipf \np{377}{92}{252}.}
\ref{Weisbergerb} {W.I.Weisberger \cmp{112}{87}{633}.}
\end{putreferences}
\newpage
\bye